\documentclass[a4paper,12pt]{article}

\pdfoutput=1
\usepackage{amsmath}
\usepackage{amssymb}
\usepackage{amsfonts}
\usepackage{mathrsfs}
\usepackage{bbm}
\usepackage{graphicx,subfigure,booktabs}
\usepackage[numbers,sort&compress]{natbib}
\usepackage{verbatim}
\usepackage{color}
\usepackage{ulem}
\usepackage{setspace}
\usepackage{multirow}
\usepackage{url}

\usepackage[utf8]{inputenc}

\usepackage[colorlinks,
linkcolor=black,
filecolor=black,
anchorcolor=black,
urlcolor=black,
citecolor=blue,
bookmarks=false,
]{hyperref}

\usepackage[hyphenbreaks]{breakurl}

\usepackage{fancyhdr}

\numberwithin{equation}{section}

\newlength{\dinwidth}
\newlength{\dinmargin}
\allowdisplaybreaks

\begin{document}

%%%%%%%%%%%%%%%%%%%%%%%%%%%%%%%%%%%%%%%%%%%%%%%%%%
%\begin{abstract}
%abs
%\end{abstract}
%%%%%%%%%%%%%%%%%%%%%%%%%%%%%%%%%%%%%%%%%%%%%%%%%%

%\maketitle

\title{Explaining anomalies of $B$-physics, muon $g-2$ and $W$ mass in $R$-parity violating MSSM with seesaw mechanism}

\author{
Min-Di Zheng${}^{a}$\footnote{zhengmd5@mail.sysu.edu.cn}\,\,,
Feng-Zhi Chen${}^{a,b}$\footnote{chenfzh25@mail.sysu.edu.cn}\,\,,
and 
Hong-Hao Zhang${}^{a}$\footnote{zhh98@mail.sysu.edu.cn}\\[12pt]
\small ${}^{a}$ School of Physics, Sun Yat-Sen University, Guangzhou 510275, China \\[-0.2cm]
\small ${}^{b}$ Key Laboratory of Quark and Lepton Physics~(MOE),\\[-0.2cm]
\small Central China Normal University, Wuhan 430079,China
}

\date{}
\maketitle

%%%%%%%%%%%%%%%%%%%%%%%%%%%%%%%%%%%%%%%%%%%%%%%%%%%%%%%%%%%%%%%%%%%%%%%%
\begin{abstract}
The recent experimental results including $R_{K^{(\ast)}}$, $R_{D^{(\ast)}}$, $(g-2)_\mu$ and $W$ mass show deviations from the standard model (SM) predictions, implying the clues of new physics (NP). In this work, we investigate the explanations of these anomalies in the $R$-parity violating minimal supersymmetric standard model (RPV-MSSM) extended with the inverse seesaw mechanism. The non-unitarity extent $\eta_{ee}$ and the loop corrections from the interaction $\lambda'\hat L \hat Q \hat D$ are utilized to raise the prediction of $W$ mass through muon decays. We also find that the interaction $\lambda'\hat L \hat Q \hat D$ involved with right-handed (RH)/singlet (s)neutrinos can explain the $R_{K^{(\ast)}}$ and $R_{D^{(\ast)}}$ anomalies simultaneously when considering nonzero $\lambda'_{1jk}$. 
For nonzero $\lambda'_{2jk}$, this model fulfills the whole $b\to s\ell^+\ell^-$ fit but cannot be accordant with $R_{D^{(\ast)}}$ measurements. The explanations in both cases are also favored by $(g-2)_\mu$ data, neutrino oscillation data and the relevant constraints we scrutinized. Furthermore, this model framework can be tested in future experiments covering, e.g., the predicted lepton flavor violations (LFV) at Belle II and the Future Circular Collider with $e^+e^-$ beams (FCC-ee), as well as the heavy neutrinos at future colliders.   
\end{abstract}
%%%%%%%%%%%%%%%%%%%%%%%%%%%%%%%%%%%%%%%%%%%%%%%%%%%%%%%%%%%%%%%%%%%%%%
\newpage

\section{Introduction}\label{sec:intro}
Recently, the experimental measurements implying the lepton flavor universality violation (LFUV) within the semileptonic decays of $B$-meson show striking results. The measurement of the observable $R_{K} = {{\cal B}(B \rightarrow K \mu^+ \mu^-)}/{{\cal B}(B \rightarrow K e^+ e^-)}$, reported by the LHCb collaboration~\cite{LHCb:2021trn} with the value $R_K=0.846^{+0.042~+0.013}_{-0.039~-0.012}$ in $q^2$ bin $[1.1,6]$ ${\rm GeV}^2$, deviates from the SM prediction by $3.1 \sigma$. The relevant $R_{K^\ast}$ measurements show  deviations larger than $2\sigma$~\cite{Aaij:2017vbb} from the SM. Besides, some discrepancies larger than $2\sigma$ are also reported in several measurements of $b\to s\mu^+\mu^-$ processes, including $P'_5$~\cite{Aaij:2020nrf}, the branching ratios ${\cal B}(B_s\to \phi\mu^+\mu^-)$~\cite{LHCb:2021zwz}, ${\cal B}(B_s\rightarrow \mu^+\mu^-)$~\cite{LHCb:2021awg,LHCb:2021vsc,Sirunyan:2019xdu}, etc. As for the charged-current decays of $B$-meson, the combined result of the experimental observables $R_{D^{(\ast)}}={\cal B}(B \to D^{(\ast)} \tau \nu) / {\cal B}(B \to D^{(\ast)} \ell \nu)$ ($\ell=e,\mu$), given by the Heavy Flavor Averaging Group (HFLAV), shows  currently $3.3\sigma$ away from the SM prediction with a  relative correlation $-0.38$~\cite{HFLAV:2022pwe,HFLAV:2019otj}. Thus, the $b\to s\ell^+\ell^-$  anomalies, especially the striking $R_K$ deviation along with  $R_{D^{(*)}}$ deviation, all imply the NP with LFUV in $B$-physics. 

Apart from the $B$-physics anomalies, clues of NP can also be probed in the precision measurements. The latest result of the muon anomalous magnetic moment $a_\mu=(g-2)_\mu$ in the E989 experiment is reported by the Fermilab National Accelerator Laboratory (FNAL) as $a_\mu^{\rm FNAL}=116592040(54)\times10^{-11}$~\cite{Muong-2:2021ojo}, which agrees with the previous result from the E821 experiment at Brookhaven National Laboratory (BNL), $a_\mu^{\rm BNL}=116592080(63)\times10^{-11}$~\cite{Muong-2:2006rrc}, but is $3.3\sigma$ away from the SM prediction $a_\mu^{\rm SM}=116591810(43)\times10^{-11}$~\cite{Aoyama:2020ynm}. Combining the two experimental results  yields $4.2\sigma$ deviation from the SM\footnote{Recent lattice calculations for the hadronic vacuum polarization~\cite{Borsanyi:2020mff,Alexandrou:2022amy,Ce:2022kxy} induce a weaker tension with this combined measurement result, than the preceding review of various SM predictions~\cite{Aoyama:2020ynm}. However, these lattice results show a tension with the $e^+e^-\to{\rm hadrons}$ data~\cite{Borsanyi:2020mff,Alexandrou:2022amy,Ce:2022kxy,Colangelo:2022vok}. Readers can see Refs.~\cite{Crivellin:2020zul,Keshavarzi:2020bfy,deRafael:2020uif,Malaescu:2020zuc,Colangelo:2020lcg} for relevant discussions.}.  
Intriguingly and very recently, another anomaly has been revealed in high precision measurement on the $W$-boson mass by the Collider Detector at Fermilab (CDF) collaboration at Tevatron. The measured value is $m_W^{\rm CDF-II}=80.4335\pm0.0094~{\rm GeV}$~\cite{CDF:2022hxs}, which shows $7\sigma$ deviation from the SM prediction $m_W^{\rm SM}=80.357\pm0.006~{\rm GeV}$~\cite{Awramik:2003rn}. This measurement, if confirmed by other experiments in the future, will profoundly change the trend of NP searches.

Combining this astonishing $m_W^{\rm CDF-II}$ with the $B$-physics anomalies and $(g-2)_\mu$ data, particular NP features are competitive for explanations. In this work, we focus on the supersymmetric (SUSY) models extended by the $R$-parity violation. As one knows, this framework with the superpotential term $\lambda'\hat L \hat Q \hat D$ among the RPV terms, can provide explanations for $B$-physics anomalies in the $b\to s\ell^+\ell^-$ or/and the $b\to c\tau\nu$ processes (see, e.g., Refs.~\cite{Altmannshofer:2017poe,Deshpand:2016cpw,Trifinopoulos:2018rna,Altmannshofer:2020axr}). In Ref.~\cite{Zheng:2021wnu}, two of us provide RPV-MSSM extended by the inverse seesaw mechanism (named RPV-MSSMIS) firstly to explain the $b\to s\ell^+\ell^-$ anomalies through the RPV-interactions including RH/singlet (s)neutrinos and explain the $(g-2)_\mu$ data through the chirality-flipping. The explanation also fulfills the recent neutrino-oscillation data. Here we ask this question: can this model further accommodate the $R_{D^{(\ast)}}$ measurements and $m_W^{\rm CDF-II}$?
The simultaneous explanations of the $R_{K^{(\ast)}}$ and $R_{D^{(\ast)}}$ anomalies in RPV-MSSM have been studied recently in Refs.~\cite{Trifinopoulos:2018rna,Altmannshofer:2020axr,Hu:2020yvs,BhupalDev:2021ipu}, while with the experiment data updated, it is straightforward to reconsider this topic in RPV-MSSMIS. Besides, this model has several distinctive effects on some observables, e.g., the branching fractions of neutrino-produced processes, the extraction of the Fermi constant $G_\mu$, and the prediction of $W$ mass.  
In Ref.~\cite{Zheng:2022irz}, we show that the loop-corrections to the vertex $W\ell\nu$ from RPV-interactions can raise the $m_W$ prediction. Interestingly, the engaging of RH neutrinos can also achieve this purpose~\cite{Blennow:2022yfm,Arias-Aragon:2022ats}, although both ways are constrained by the relevant experiments, e.g., the purely leptonic decays of $Z$-boson, pion, and kaon. 
One can see that the two ways are both included in RPV-MSSMIS, so it is promising to explain (some of) $B$-physics anomalies, $(g-2)_\mu$ data, and $m_W$ shift simultaneously.

Our paper is organized as follows. The model and the theoretical calculations are in section~\ref{sec:anomaly}. Then we scrutinize the relevant constraints in section~\ref{sec:constraints}, which is followed by the numerical results and discussions in section~\ref{sec:num}. The conclusions are finally made in section~\ref{sec:conclusion}. 

\section{The anomaly explanations in RPV-MSSMIS}\label{sec:anomaly}

In this section, we begin to investigate the NP effects on the flavor anomalies, i.e., $b\to s\ell^+\ell^-$ anomalies, $R_{D^{(*)}}$ deviations, and the $(g-2)_\mu$ problem, as well as the new measurement of $m_W^{\rm CDF-II}$, in the framework of RPV-MSSMIS.

\subsection{RPV-MSSMIS framework}
\label{sec:model}
First we make some introductions to RPV-MSSMIS~\cite{Zheng:2021wnu}. The superpotential and the soft SUSY breaking Lagrangian are respectively given by
\begin{align}\label{eq:MSSMIS-RPV}
{\cal W} =& {\cal W}_{\rm MSSM} 
+ Y_\nu^{ij} \hat R_i \hat L_j \hat H_u + M_R^{ij} \hat R_i \hat S_j + \frac{1}{2} \mu_S^{ij} \hat S_i \hat S_j
+ \lambda'_{ijk} \hat L_i \hat Q_j \hat D_k,   \notag\\
-{\cal L}^{\rm soft}=&-{\cal L}^{\rm soft}_{\rm MSSM}
+(m^2_{\tilde{R}})_{ij} \tilde{R}^\ast_i\tilde{R}_j
+(m^2_{\tilde{S}})_{ij} \tilde{S}^\ast_i\tilde{S}_j \notag\\
&+(A_\nu Y_{\nu})_{ij} \tilde{R}^\ast_i\tilde{L}_jH_u
+B_{M_R}^{ij} \tilde{R}^\ast_i\tilde{S}_j
+\frac{1}{2} B_{\mu_S}^{ij} \tilde{S}_i\tilde{S}_j ,
\end{align} 
where the MSSM parts, ${\cal W}_{\rm MSSM}$ and ${\cal L}^{\rm soft}_{\rm MSSM}$, can be referred to Refs~\cite{Rosiek:1989rs,Rosiek:1995kg}, and the neutrino sector consists of pairs of SM singlet superfields, $\hat R_i$ and $\hat S_i$. 
The generation indices $i,j,k=1,2,3$ while the colour ones are omitted, and
all the repeated indices are defaulted to be summed over unless otherwise stated. Besides, squarks/sleptons are denoted by ``$\tilde{\ }$'' here. 
The neutral scalar fields of the two Higgs doublet superfields, $\hat H_u=(\hat H^+_u,\hat H^0_u)^T$ and $\hat H_d=(\hat H^0_d,\hat H^-_d)^T$, acquire the non-zero vacuum expectation value, i.e., $\langle H^0_u \rangle=v_u$ and $\langle H^0_d \rangle=v_d$, respectively, and their mixing is expressed by $\tan\beta=v_u/v_d$.

The additional neutrino sector in the superpotential ${\cal W}$ provides the neutrino mass spectrum at the tree level, and the $9\times9$ mass matrix in the basis $(\nu, R, S)$ is
\begin{align}\label{eq:mnu}
{\cal M}_{\nu} = \left( 
\begin{array}{ccc}
0 &m_D^{T}  &0\\ 
m_D  &0 &M_R\\ 
0 &M_{R}^{T} &\mu_S\end{array} 
\right), 
\end{align}
where the Dirac mass matrix $m_D = \frac{1}{ \sqrt{2}} v_u Y_\nu^T$. Then ${\cal M}_{\nu}$ can be diagonalized through ${\cal M}^{\text{diag}}_{\nu}={\cal V} {\cal M}_{\nu} {\cal V}^T$. As to the sneutrino mass square matrix in the basis $(\tilde{\nu}^{\cal I(R)}_{L},\tilde{R}^{\cal I(R)},\tilde{S}^{\cal I(R)})$, it is expressed as
\begin{align}\label{eq:mSnu}
{\cal M}_{\tilde{\nu}^{\cal I(R)}}^2  =& 
\left(\begin{array}{ccc} 
m^2_{\tilde{L}'} & (A_\nu -\mu\cot\beta) m_D^T & m_D^T M_R \\
(A_\nu -\mu\cot\beta) m_D & m^2_{\tilde{R}}+M_RM_R^{T}+m_Dm_D^{T} & 
\pm M_R\mu_S + B_{M_R} \\
M_R^T m_D & \pm \mu_S M_R^{T} + B_{M_R}^T 
& m^2_{\tilde S}+ \mu_S^2+M_R^TM_R \pm B_{\mu_S}
\end{array}\right)  \notag\\
\approx &
\left(\begin{array}{ccc} 
m^2_{\tilde{L}'} & (A_\nu -\mu\cot\beta) m_D^T & m_D^T M_R \\
(A_\nu -\mu\cot\beta) m_D & m^2_{\tilde{R}}+M_RM_R^{T}+m_Dm_D^{T} & B_{M_R} \\
M_R^T m_D & B_{M_R}^T 
& m^2_{\tilde S}+M_R^T M_R
\end{array}\right),
\end{align}
where ``$\pm$'', as well as ``${\cal R(I)}$'', denotes the CP-even (odd), and the mass square $m^2_{\tilde{L}'}=m^2_{\tilde{L}}+\frac{1}{2} m^2_Z \cos 2\beta+m_Dm_D^T$ is regarded as a whole input with $m^2_{\tilde{L}}$ being the soft mass square of $\tilde{L}$. Given that the value of $\mu_S$ is tiny and $B_{\mu_S}$ can be also relatively small~\cite{BhupalDev:2012ru}, we have obtained the approximate  result in Eq.~\eqref{eq:mSnu}, which induces the CP-even and CP-odd masses to be nearly the same. With this approximation, the mixing matrices $\tilde{\cal V}^{\cal I(R)}$, which diagonalize sneutrino mass square matrices through $\tilde{\cal V}^{\cal I(R)} {\cal M}_{\tilde{\nu}^{\cal I(R)}}^2 \tilde{\cal V}^{\cal I(R) \dagger}={({\cal M}_{\tilde{\nu}^{\cal I(R)}}^2)}^{\text{diag}}$, can also be regarded as the same whether CP is even or odd. In the rest of the paper, we can replace $\tilde{\cal V}^{\cal R}$ and the physical mass $m_{\tilde{\nu}^{\cal R}}$ with the corresponding notations of $\tilde{\cal V}^{\cal I}$ and $m_{\tilde{\nu}^{\cal I}}$.

Afterwards, we introduce the trilinear RPV interaction in this model. With the superpotential term $\lambda'_{ijk} \hat L_i \hat Q_j \hat D_k$, the relevant Lagrangian in the context of mass eigenstates for the down-type quarks and the charged leptons is given by
\begin{align}\label{eq:RPVlagphys}
{\cal L}_{\text{LQD}} =& \lambda'^{\cal I}_{vjk} \tilde{\nu}_{v} \bar{d}_{Rk} d_{Lj} 
+\lambda'^{\cal N}_{vjk} \big(\tilde{d}_{Lj} \bar{d}_{Rk} \nu_{v} + \tilde{d}_{Rk}^\ast \bar{\nu}_{v}^c d_{Lj} \big)  \notag\\
&-\tilde{\lambda}'_{ilk} \big(\tilde{l}_{Li} \bar{d}_{Rk} u_{Ll} + \tilde{u}_{Ll} \bar{d}_{Rk} l_{Li} + \tilde{d}_{Rk}^\ast \bar{l}_{Li}^c u_{Ll}\big) + {\rm h.c.},
\end{align}
where ``$c$'' indicates the charge conjugated fermions, and the fields $\tilde{\nu}_{L}$, $\nu_{L}$, and $u_{L}$ (aligned with $\tilde{u}_L$) in the flavor basis have been rotated into mass eigenstates by the mixing matrices $\tilde{\cal V}^{\cal I}$, ${\cal V}$, and $K$, respectively. Concretely, the indices $v=1,2,\dots 9$ denote the generation of the physical (s)neutrinos, and the three $\lambda'$ couplings are deduced as $\lambda'^{\cal I}_{vjk}=\lambda'_{ijk} \tilde{\cal V}^{\cal I \ast}_{vi}$, $\lambda'^{\cal N}_{vjk}=\lambda'_{ijk} {\cal V}_{vi}$ and 
$\tilde{\lambda}'_{ilk}=\lambda'_{ijk} K^{\ast}_{lj}$. In the following, the interaction (diagram) involving these $\lambda'$ couplings is called the $\lambda'$ interaction (diagram).

By the end of this section, we mention the chargino mass matrix from the MSSM sector, which is~\cite{Rosiek:1995kg}
\begin{align}
{\cal M}_{\chi^\pm}=\left(
\begin{array}{cc}
 M_2 & \frac{e v_u}{\sqrt{2} \sin\theta_W} \\
 \frac{e v_d}{\sqrt{2} \sin\theta_W} & \mu  \\
\end{array}
\right),
\end{align}
with $M_2$ being the wino mass, $\mu$ being the Higgsino mass in the flavor basis, and $\theta_W$ being the Weinberg angle.
The mixing matrices $V$ and $U$ diagonalize ${\cal M}_{\chi^\pm}$ by $U^{\ast} {\cal M}_{\chi^\pm} V^{\dagger}= {\cal M}_{\chi^\pm}^{\text{diag}}$. For a review on the neutralino matrix, readers are also referred to Ref.~\cite{Rosiek:1995kg}.

\subsection{$b\to s\ell^+\ell^-$ anomalies}
\label{sec:bsll}
To study the NP effects on the $b\to s\ell^+\ell^-$ process, the relevant Lagrangian of the low energy effective field theory can be represented as
\begin{equation}
{\cal L}_{\rm eff}^{bs\ell\ell} = \frac{4 G_F}{\sqrt{2}} \eta_t \sum_{i}C_i{\cal O}_i + {\rm h.c.},
\end{equation}
where $\eta_t \equiv K_{tb} K_{ts}^{\ast}$ is the Cabibbo–Kobayashi–Maskawa (CKM) factor. The most favored operators to explain the $b\to s\ell^+\ell^-$ anomalies are
\begin{align}
{\cal O}_9 = \frac{e^2}{16\pi^2}(\bar{s}\gamma_\mu P_{L} b)(\bar{\ell}\gamma^\mu \ell), \quad
{\cal O}_{10} = \frac{e^2}{16\pi^2}(\bar{s}\gamma_\mu P_{L} b)(\bar{\ell}\gamma^\mu \gamma_5 \ell),
\end{align}
where $P_{L}=(1-\gamma_5)/2$ is the left-handed (LH) chirality projector, and the Wilson coefficients $C_{9(10)} = C_{9(10)}^{\rm SM} + C_{9(10)}^{\rm NP}$. Then with the results of the model-independent global fit~\cite{Alok:2019ufo,Alda:2020okk,Carvunis:2021jga,Geng:2021nhg,Li:2021toq,Angelescu:2021lln,Altmannshofer:2021qrr,Cornella:2021sby,Kriewald:2021hfc,Isidori:2021vtc,Alguero:2021anc,Hurth:2021nsi,Perez:2021ddi,Alda:2021ruz,Bause:2021cna}, RPV-MSSMIS can provide box contributions that are restricted to the $\mu\mu(ee)$ channel, i.e., $C_{9\mu}^{\rm NP}=-C_{10\mu}^{\rm NP}<0$ or $C_{9e}^{\rm NP}=-C_{10e}^{\rm NP}>0$ to explain the $R_{K^{(\ast)}}$ anomaly. 
Especially, the case $C_{9\mu}^{\rm NP}=-C_{10\mu}^{\rm NP}<0$ can be further in accord with the deviations of $b\to s\mu^+\mu^-$ measurements from the SM predictions, e.g., measurements of $B_s \to \mu^+\mu^-$~\cite{LHCb:2021awg,LHCb:2021vsc,Sirunyan:2019xdu} and $B_s\rightarrow \phi\mu^+\mu^-$~\cite{LHCb:2021zwz,LHCb:2015wdu}.
 
In RPV-MSSMIS, we set sleptons and winos with masses of several $10^2$~GeV while the masses of all colored sparticles as well as the heavy neutrinos are around TeV or above, and all the model parameters are set at the scale $\mu_{\rm NP}=0.5$~TeV. To eliminate the unfavored contributions to $C_{9(10)}^{\prime\rm NP}$ (the operator ${\cal O}_{9(10)}'$ is given by replacing $P_L$ with $P_R$ in ${\cal O}_{9(10)}$) which can emerge at the tree level, we assume the coupling $\lambda'_{ijk}$ is non-negligible only for the single value $k$ at $\mu_{\rm NP}$ scale~\cite{Zheng:2021wnu}. Then the dominant contributions to $C_{9\ell}^{\rm NP}=-C_{10\ell}^{\rm NP}$ in RPV-MSSMIS are given by
\begin{align}\label{eq:Wcsbsll}
C_{9\ell}^{\tilde{\nu}-\chi^\pm}=\lambda'^{\cal I}_{v3k} \lambda'^{\cal I \ast}_{v'2k} (g_2 V^{\ast}_{m1} \tilde{\cal V}^{\cal I}_{v\ell}-V^{\ast}_{m2} Y^{\cal I}_{\ell v}) 
(g_2 V_{m1} \tilde{\cal V}^{\cal I}_{v'\ell}-V_{m2} Y^{\cal I}_{\ell v'}) D_2[m_{\tilde{\nu}^{\cal I}_{v}},m_{\tilde{\nu}^{\cal I}_{v'}},m_{\chi^\pm_m},m_{d_k}],
\end{align}
where the formula of Passarino-Veltman function~\cite{Passarino:1978jh} $D_2$ is collected in appendix~\ref{app:bsllCbox}, and the coupling $Y^{\cal I}_{\ell v} \equiv {(Y_\nu)}_{j\ell} \tilde{\cal V}^{\cal I \ast}_{v(j+3)}$. One can see that the light sneutrinos and winos will help provide considerable NP effects. The Wilson coefficient of Eq.~\eqref{eq:Wcsbsll} is proportional to the product $\lambda'_{2(1)3k}\lambda'^\ast_{2(1)2k}$, which is related to the $\mu\mu(ee)$-channel contribution, assuming no flavor transitions within sneutrino sectors. In appendix~\ref{app:bsllCbox}, we provide the whole list of formulas  from the scrutinized one-loop box diagrams of the $b\to s\ell^+\ell^-$ process, which are adopted in the numerical calculations. 
Under the assumption of single value $k$, only the LH-quark-vector-current contributions $C_{9(10)\ell}^{\rm NP}(\mu_{\rm NP})$ exist, and the RH-quark-vector-current contributions $C_{9(10)\ell}^{\prime\rm NP}(\mu_{\rm NP})$ vanish. 
Due to the approximate conservation of (axial-)vector currents, there is $C_{9(10)\ell}^{\rm NP}(\mu_b)=C_{9(10)\ell}^{\rm NP}(\mu_{\rm NP})$, and $ C_{9(10)\ell}^{\prime\rm NP}(\mu_b)$ still vanish when the scale runs down to $\mu_b=m_b$ through QCD renormalization. Then we can adopt the results at $\mu_b$ from the model-independent global fit in Ref.~\cite{Altmannshofer:2021qrr} to constrain the model inputs. The fit results show $C_{9e}^{\rm NP}=-C_{10e}^{\rm NP}=0.37\pm 0.10$ as the best fit for the $R_{K^{(\ast)}}$ explanations through NP in the $ee$ channel. As for the $\mu\mu$ channel, $C_{9\mu}^{\rm NP}=-C_{10\mu}^{\rm NP}=-0.35\pm 0.08$ and $C_{9\mu}^{\rm NP}=-C_{10\mu}^{\rm NP}=-0.39\pm 0.07$ are utilized to explain $R_{K^{(\ast)}}$ and the whole $b\to s\ell^+\ell^-$ anomaly, respectively. In this work, we restrict $k=3$ for a benchmark and consider one of $(\lambda'_{1j3},\lambda'_{2j3})$ nonzero at a time, i.e., $\lambda'_{2(1)j3}=0$ for Case A(B).

\subsection{$R_{D^{(\ast)}}$ anomalies}
\label{sec:RD}
Next, we turn to $R_{D^{(*)}}$, implying the LFUV anomalies in the charged current. For the generic charged current process $d_j \to u_n l_l \nu$, the effective Lagrangian is
\begin{align}
{\cal L}_{\rm eff}^{dul\nu} = -\frac{4G_F}{\sqrt{2}} K_{nj}
({\cal V}^T_{li} + {\cal V}^T_{i' i} C_{njli'}) \bar{u}_n \gamma_{\mu} P_L d_j \bar{l}_l \gamma^{\mu} P_L \nu_i + {\rm h.c.}, 
\end{align}
where the first term in the bracket following the CKM element $K_{nj}$ gives the SM contribution combined with the neutrino-generation mixing, and the second term is related to the $\lambda'$ interactions with the function
\begin{align}
C_{njli'}
= \frac{ \tilde{\lambda}'^{\ast}_{ln3} \lambda'_{i' j3} }{4 \sqrt{2} G_F K_{nj} m_{\tilde{b}_{R}}^2}.
\end{align} 
It is useful to define the ratio
\begin{align}\label{eq:RDratio}
R_{njl} \equiv& \frac{{\cal B}(d_j \to u_n l_l \nu)_{\rm SM+NP}}{{\cal B}(d_j \to u_n l_l \nu)_{\rm SM}}
=\frac{\sum\limits_{i=1}^{3} \left|K_{nj}\right|^2 \left|{\cal V}^T_{li}+ {\cal V}^T_{i' i} C_{njli'} \right|^2}{\sum\limits_{i=1}^{3} \left|K_{nj} \delta_{li} \right|^2} \notag \\
&=\sum_{i,i'=1}^3 \left|{\cal V}^T_{i'i}\right|^2 \left|\delta_{li'}+ C_{njli'} \right|^2
\approx \sum_{i'=1}^3 \left|\delta_{li'}+ C_{njli'} \right|^2.
\end{align}
Thus, it approximates the ratio in ordinary RPV-MSSM (see Eq.~(24) in Ref.~\cite{Hu:2020yvs}) under nearly the unitarity bound of ${\cal V}^T_{3\times 3}$. Then we get the ratio,
\begin{align}\label{eq:RDratio2}
\frac{R_D}{R_D^{\rm SM}}=\frac{R_{D^\ast}}{R_{D^\ast}^{\rm SM}}= \frac{2R_{233}}{R_{232} + R_{231}} = 
\begin{cases}
2\frac{
\bigl|\tilde{\lambda}'^{\ast}_{323}\lambda'_{133}\bigr|^2+\bigl|4\sqrt{2}G_F K_{23} m_{\tilde{b}_{R}}^2+\tilde{\lambda}'^{\ast}_{323}\lambda'_{333}\bigr|^2
}{
\bigl|\tilde{\lambda}'^{\ast}_{123}\lambda'_{333}\bigr|^2+\bigl|4\sqrt{2}G_F K_{23} m_{\tilde{b}_{R}}^2+\tilde{\lambda}'^{\ast}_{123}\lambda'_{133}\bigr|^2
},& \text{in Case A}\\[5mm]
2\frac{
\bigl|\tilde{\lambda}'^{\ast}_{323}\lambda'_{233}\bigr|^2+\bigl|4\sqrt{2}G_F K_{23} m_{\tilde{b}_{R}}^2+\tilde{\lambda}'^{\ast}_{323}\lambda'_{333}\bigr|^2
}{
\bigl|\tilde{\lambda}'^{\ast}_{223}\lambda'_{333}\bigr|^2+\bigl|4\sqrt{2}G_F K_{23} m_{\tilde{b}_{R}}^2+\tilde{\lambda}'^{\ast}_{223}\lambda'_{233}\bigr|^2
},& \text{in Case B}
\end{cases},
\end{align}
which is named as ${\cal R}^{\rm NP/SM}_{D^{(\ast)}}$ in this work. 
We utilize the $R_{D^{(\ast)}}$ world average with the so-called correlation $\rho_{D^{\ast\ast}}$, which denotes the $B\to D^{\ast\ast} \ell \bar{\nu}_\ell$ correlation structure across $R_D$ and $R_{D^\ast}$ measurements, as zero~\cite{Bernlochner:2021vlv}. This result is similar to the one of HFLAV~\cite{HFLAV:2022pwe,HFLAV:2019otj}. Then in this model, we get the fit value 
${\cal R}^{\rm NP/SM}_{D^{(\ast)}}=1.140 \pm 0.045$.

\subsection{$m_W$ shift}
\label{sec:mW}
Here we discuss the NP contributions to the $W$-boson mass  from the $W\ell\nu$-vertex loop corrections involving $\lambda'$ couplings and the non-unitarity of ${\cal V}^T_{3\times 3}$ in this model.

The non-unitarity of ${\cal V}^T_{3\times 3}$ can be shown in
\begin{align}\label{eq:eta}
\left({\cal V}^T_{3\times 3}\right)_{ij}=(\delta_{ik} + \eta_{ik}) {\cal U}_{kj},
\end{align}  
where ${\cal U}$ is unitary, and the Hermitian $\eta$ describes the non-unitarity extent. In the inverse seesaw framework, one can figure out $\eta\approx -\frac{1}{2} m_D^\dagger (M_R^\ast)^{-1} (M_R^T)^{-1} m_D$.

Then we turn to the $Wl\nu$-vertex expressed in the Lagrangian,
\begin{align}
{\cal L}^{Wl\nu}_{\rm eff}=\frac{e}{\sqrt{2}\sin\theta_W} \bar{l_l} \gamma^\mu P_L ({\cal V}^T_{li}+h_{li}) \nu_i W^{-}_\mu + {\rm h.c.},
\end{align}
where the first part in the bracket shows the SM contribution combined with the neutrino-generation mixing. The term ${\cal V}^T_{li}$ can be replaced by $\delta_{li} + \eta_{li}$ by dropping out of the matrix ${\cal U}$ due to the limit of vanishing $m_{\nu_i}$~\cite{Bryman:2021teu}.
The one-loop correction part, $h_{li}$, is dominated by the $\lambda'$ contribution $h'_{li}$. As the analogy to the formula in Ref.~\cite{Arnan:2019olv}, this contribution is given by~\cite{Zheng:2022irz}
\begin{align}\label{eq:hprime}
h'_{li} = -\frac{3}{64\pi^2} x_{\tilde{b}_R} f_W(x_{\tilde{b}_R}) \tilde{\lambda}'^{\ast}_{l33} \tilde{\lambda}'_{i33},
\end{align}
where $x_{\tilde{b}_R} \equiv m^2_t/m^2_{\tilde{b}_R}$ and the loop function $f_W(x) \equiv \frac{1}{x-1} + \frac{(x-2) \log x}{(x-1)^2}$, and other minor parts including the ones proportional to $\eta h'$ product are eliminated. This dominant part is from the $u_id_i\tilde{b}_R$ loop diagram, 
in which the engaging top quark with large mass and couplings $\tilde{\lambda}'_{l(i)33}$ make $h'_{li}$ dominant.

Since $h'$ and $\eta$ can contribute to the $W\ell\nu$-vertex at the same level, they both affect the muon decay and induce (see similar formulas in Refs.~\cite{Bryman:2021teu,Blennow:2022yfm})
\begin{align}\label{eq:Gmu}
G_\mu=G_F(1+\eta_{\ell\ell}+h'_{\ell\ell}),
\end{align}
where $G_\mu$ is the Fermi constant extracted from the muon lifetime, while $G_F$ corresponds to the one in the SM. When $m_D$ and $M_R$ are set diagonal, matrix $\eta_{3\times3}$ can be diagonal. Since we consider the index $i$ in the $\lambda'_{ijk}$ as $1$ or $2$ at a time, the submatrix $h'_{2\times2}$ (without $\tau$) is also diagonal. Here the $W\ell\nu_\tau$-effect from pure NP is omitted because it has no interference with the SM contribution, and thus we only keep $h'_{\ell\ell}$ without $h'_{\ell\tau}$ in Eq.~\eqref{eq:Gmu}. 
Then the prediction of $m_W$ is given by
\begin{align}\label{eq:mwmass}
\frac{m^2_W}{m^2_Z}=\frac{1}{2}+\sqrt{\frac{1}{4}-\frac{\pi \alpha}{\sqrt{2} G_\mu m^2_Z} (1+\eta_{\ell\ell}+\Delta r_0+h'_{\ell\ell})},
\end{align}
where $\Delta r_0$ represents the loop corrections from the SM and pure MSSM. The MSSM part contributes mainly to the $W$ self-energy that is not considered in this work by setting sufficiently heavy up-type squarks. One can see that the negative values of $h'_{\ell\ell}$ or/and $\eta_{\ell\ell}$ can raise the prediction of $m_W$ to approach the CDF-II measurement. Besides, these two variants are related to the CKM elements extraction described as 
\begin{align}\label{eq:cabbibo}
K^\beta_{ud}=K_{ud}(1-\eta_{\mu\mu}-h'_{\mu\mu}),
\end{align}
where $K^\beta_{ud}$ is extracted from beta decay. The terms $\eta_{ee}$ and $h'_{ee}$ enter both the $We\nu$ vertex and $G_\mu$, which induces the canceling in Eq.~\eqref{eq:cabbibo}. With the extraction of $K_{ud}$ and the analogous one of $K_{us}$ from kaon decay, it is found that the positive values of $\eta_{\mu\mu}+h'_{\mu\mu}$ are needed to alleviate the Cabbibo anomaly showing around $3\sigma$ tension, and besides, the simple RH-neutrino extension cannot fully explain the data~\cite{Coutinho:2019aiy}. Given that the inverse seesaw framework provides negative $\eta_{\ell\ell}$ and $h'_{\ell\ell}$ is also negative commonly, we can set model parameters to make both $|\eta_{\mu\mu}|$ and $|h'_{\mu\mu}|$ sufficiently small to avoid worse tension. 

Thus, in this model, we utilize the $We\nu_e$-vertex loop correction $h'_{ee}(m_{\tilde{b}_R},\tilde{\lambda}'_{133})$ and the non-unitarity extent $\eta_{ee}(m_D^{11},M_R^{11})$ to explain the CDF-II measurement. 
In Case A as defined in section~\ref{sec:bsll}, we need relative large $|\eta_{ee}+h'_{ee}| \gtrsim 2\times 10^{-3}$ favored to raise $m_W$ with $|\eta_{\mu\mu}| \sim 10^{-4}$; while in Case B, there is $|\eta_{ee}| \gtrsim 2\times 10^{-3}$ with both $|\eta_{\mu\mu}|$ and $|h'_{\mu\mu}|$ around $10^{-4}$. It is worth mentioning that other measurements of $\sin \theta_W$, which are in tension with $m^{\rm CDF-II}_W$, also constrain $\eta_{ee}$~\cite{Antusch:2014woa,Antusch:2016brq,Fernandez-Martinez:2016lgt}, e.g., $|\eta_{ee}|<1.3\times 10^{-3}$~\cite{Fernandez-Martinez:2016lgt}. In this work, this bound is not considered as the essential one.

\subsection{$(g-2)_\mu$}
At the end of section~\ref{sec:anomaly}, we will mention the NP contributions to $a_\mu$ in RPV-MSSMIS. 
Here we mainly utilize the one-loop chargino and neutralino diagrams to explain, and these contributions to $a_\ell$ are given by~\cite{Zheng:2021wnu}
\begin{align}\label{eq:aell}
&\delta a_{\ell}^{\chi^\pm}  =  \frac{m_\ell}{16\pi^2}
\left[  \frac{m_\ell}{ 6 m^2_{\tilde\nu_v}}
   \left(|c^{\ell L}_{mv}|^2+|c^{\ell R}_{mv}|^2\right) F^C_1(m^2_{\chi^\pm_m}/m^2_{\tilde{\nu}_v^{\cal I}})
   +\frac{m_{\chi^\pm_m}}{m^2_{\tilde\nu_v}}
         {\rm Re}( c^{\ell L}_{mv} c^{\ell R}_{mv}) F^C_2(m^2_{\chi^\pm_m}/m^2_{\tilde{\nu}_v^{\cal I}})
   \right],
   \notag \\
&\delta a_\ell^{\chi^0}  = \frac{m_\ell}{16\pi^2}
\left[ -\frac{m_\ell}{ 6 m^2_{\tilde{l}_i}}
  \left( |n^{\ell L}_{ni}|^2+ |n^{\ell R}_{ni}|^2 \right) F^N_1(m^2_{\chi^0_n}/m^2_{\tilde{l}_i})
+ \frac{m_{\chi^0_n}}{ m^2_{\tilde{l}_i}}
    {\rm Re}(n^{\ell L}_{ni} n^{\ell R}_{ni}) F^N_2(m^2_{\chi^0_n}/m^2_{\tilde{l}_i})  
   \right],
\end{align}
with
\begin{align}
&c^{\ell R}_{mv}  =  y_\ell U_{m2} \tilde{\cal V}^{\cal I }_{v\ell}, \quad
c^{\ell L}_{mv}  = - g_2 V_{m1} \tilde{\cal V}^{\cal I }_{v\ell}+V_{m2} Y^{\cal I}_{\ell v}; \notag\\
&n^{\ell R}_{ni}  = \sqrt{2} g_1 N_{n1} \delta_{i(\ell+3)} + y_{\ell} N_{n3} \delta_{i\ell},
\quad
n^{\ell L}_{ni}  = \frac{1}{\sqrt{2}} \left (g_2 N_{n2} + g_1 N_{n1} \right) \delta_{i\ell}
-y_\ell N_{n3} \delta_{i(\ell+3)},
\end{align}
where $m_{\chi^0_n}$ is the neutrino mass after the diagonalization $N {\cal M}_{\chi^0} N^T={\cal M}_{\chi^0}^{\text{diag}}$, and factor functions are defined as 
\begin{align}
&F^C_1(x) \equiv \frac{1}{(1-x)^4}\left( 2+ 3x - 6x^2 + x^3 +6x\log x\right), \notag\\
&F^C_2(x) \equiv -\frac{1}{(1-x)^3}\left( 3-4x+x^2+2\log x\right), \notag\\
&F^N_1(x) \equiv \frac{1}{(1-x)^4}\left( 1-6x+3x^2+2x^3-6x^2\log x\right),  \notag\\
&F^N_2(x) \equiv \frac{1}{(1-x)^3}\left( 1-x^2+2x\log x\right).
\end{align}
Also, the contribution from the $\lambda'$ diagrams, $\delta a_{\ell}^{\lambda'}={m^2_\ell |\lambda'_{\ell j 3}|^2}/{32\pi^2 m^2_{\tilde{b}_R}}$~\cite{Altmannshofer:2020axr}, is not dominant here. Compared with the original MSSM, the term  $V_{m2} Y^{\cal I}_{\ell v}$ in $c^{\ell L}_{mv}$ of Eq.~\eqref{eq:aell} provides extra chirality flips. With the measured deviation $\Delta a_\mu=a_\mu^{\rm{exp}}-a_\mu^{\rm{SM}}=(2.51\pm 0.59)\times 10^{-9}$~\cite{Muong-2:2021ojo}, the explanation favors $1.33 <|\delta a^{\chi^\pm}_\mu+\delta a^{\chi^0}_\mu + \delta a_{\mu}^{\lambda'}| \times 10^9 < 3.69$ at the $2\sigma$ level, contributed by the sectors of chargino, neutralino, and sleptons along with the coupling $\lambda'_{2j3}$ and the mass of ${\tilde{b}_R}$.

\section{The constraints}\label{sec:constraints}

Before the numerical explanations of the anomalies, the relevant experimental constraints should be scrutinized to determine the setting of parameter inputs.

\subsection{Direct searches} 
\label{sec:cons_direct}
First, we consider direct searches for NP particles. The no signs of SUSY particles until the end of the LHC run II, which reached around $140$ fb$^{-1}$ at the center of mass  energy of $13$ TeV, induces the stringent bounds on SUSY models. The allowed masses of colored sparticles, such as gluinos, the first-two generation squarks, stops and sbottoms have been excluded up to $1-2$ TeV scale~\cite{Aaboud:2017opj,CMS:2018qxv,ATLAS:2019gqq,ATLAS:2020xyo,ATLAS:2021fbt,CMS:2021beq,CMS:2021eha}. In this work, the masses of colored sparticles, except RH sbottoms $\tilde{b}_R$, are all set around $10$ TeV, whereas the masses of $\tilde{b}_R$, sleptons, charginos, neutralinos, charged Higgs, and the heavy neutrinos are all around $10^2-10^3$ GeV. Some recent experiments have pushed the lower limit of slepton masses over TeV scale~\cite{ATLAS:2018rns,ATLAS:2018mrn,ATLAS:2021yyr}, however,  these searches consider nonzero $\lambda$ in the superpotential $\lambda_{ijk} \hat L_i \hat L_j \hat E_k$. Given that we only consider nonzero $\lambda'$ in the model, so processes of purely leptonic decays of sleptons can be neglected, and one can focus on sleptons decaying to the lightest neutralino $\chi^0_1$ and leptons. Although the dijet resonance pairs can emerge from pair-produced sleptons through the $\lambda'$ interactions, this provides relatively weak bounds due to the large QCD background~\cite{Agashe:2022uih}. Thus, we take a compressed scenario, $m_{\chi^\pm_1}\gtrsim m_{\chi^0_1}\gtrsim 300$ GeV as well as $m_{\tilde{l}_L}>300$ GeV based on the recent searches~\cite{ATLAS:2019lff,ATLAS:2019lng}.

\subsection{Tree-level processes}

As we set RH sbottom not decoupled, the tree-level processes exchanging sbottoms will make constraints on the model parameters and these relevant processes include $B\rightarrow K^{(\ast)}\nu\bar{\nu}$, $B \to \pi \nu \bar\nu$, $K^+ \to \pi^+ \nu \bar\nu$, $D^0 \to \ell^+ \ell^-$, and $\tau \rightarrow \ell \rho^0$, as well as the charged current processes $B \to \tau \nu$, $D_s \to \tau \nu$, $\tau \to K(\pi) \nu$, and $\pi\to\ell\nu(\gamma)$. The first to be introduced are the semileptonic decays $B\rightarrow K^{(\ast)}\nu\bar{\nu}$, $B \to \pi \nu \bar\nu$, and $K^+ \to \pi^+ \nu \bar\nu$ involving $d_j\rightarrow d_m \nu_i\bar{\nu}_{i'}$. The related effective Lagrangian is defined by
\begin{align}
{\cal L}_{\rm eff}^{dd\nu\bar{\nu}} = (C^{\rm SM}_{mj} \delta_{ii'} + C^{\nu_i\bar{\nu}_{i'}}_{mj}) (\bar{d}_m \gamma_\mu P_L d_j)(\bar{\nu}_i \gamma^\mu P_L \nu_{i'}) + {\rm h.c.},
\end{align}
where the SM contribution is $C^{\rm SM}_{mj}=-\frac{\sqrt{2} G_F e^2 K_{tj}K^\ast_{tm}}{4\pi^2 \sin^2\theta_W} X(x_t)$ and the loop function $X(x_t) \equiv \frac{x_t(x_t+2)}{8(x_t-1)}+\frac{3 x_t (x_t-2)}{8(x_t-1)^2}\log(x_t)$ with $x_t \equiv m^2_t/m^2_W$~\cite{Buras:2014fpa}. The NP contributions are
\begin{align}\label{eq:Cbsnunu}
C^{\nu_i\bar{\nu}_{i'}}_{mj} =\frac{\lambda'^{\cal N}_{i'j3}\lambda'^{\cal N\ast}_{im3}}{2 m^2_{\tilde{b}_{R}}}
= \frac{{\cal V}_{i'\alpha'} {\cal V}^{\ast}_{i\alpha} \lambda'_{\alpha'j3} \lambda'^{\ast}_{\alpha m3}}{2 m^2_{\tilde{b}_{R}}}.
\end{align}
It is worth mentioning that the difference between Eq.~\eqref{eq:Cbsnunu} and the corresponding formula in the ordinary RPV-MSSM~\cite{Deshpande:2016yrv} is the neutrino-generation mixing in $\lambda'$ interactions.  With Eq.~\eqref{eq:eta}, we can further get
\begin{align}
C^{\nu_i\bar{\nu}_{i'}}_{mj} \approx \frac{{\cal U}_{\alpha'i'} {\cal U}^{\ast}_{\alpha i} \lambda'_{\alpha'j3} \lambda'^{\ast}_{\alpha m3}}{2 m^2_{\tilde{b}_{R}}}.
\end{align}
Given that ${\cal U}$ is not a diagonal-like matrix, the NP effects on this process are unique compared to RPV-MSSM.  

The experimental measurement ${\cal B}(K^+ \to \pi^+ \nu \bar\nu)_{\rm exp}=(1.7\pm1.1)\times 10^{-10}$~\cite{ParticleDataGroup:2020ssz} combined with the SM prediction ${\cal B}(K^+ \to \pi^+ \nu \bar\nu)_{\rm SM}=(9.24\pm0.83)\times 10^{-11}$~\cite{Aebischer:2018iyb} induces the strong constraint that $|\lambda'^{\cal N}_{i'2k}\lambda'^{\cal N\ast}_{i1k}|<7.4\times 10^{-4}(m_{\tilde{b}_R}/1 {\rm TeV})^2$~\cite{Hu:2020yvs}. Thus, we assume negligible $\lambda'_{i1k}$ to avoid the bound, and then the $B \to \pi \nu \bar\nu$ process will also make no bound.

Then we investigate the constraint from $B\rightarrow K^{(\ast)}\nu\bar{\nu}$. One can define the ratio,
\begin{align}
R^{\nu\bar{\nu}}_{\rm K^{(\ast)}} \equiv \frac{{\cal B}(b \rightarrow s \nu\bar{\nu})_{\rm NP+SM}}{{\cal B}(b \rightarrow s \nu\bar{\nu})_{\rm SM}} 
= \frac{\sum\limits_{i=1}^3 \left| C_{23}^{\rm SM} + C_{23}^{\nu_i \bar{\nu}_i }\right|^2 + \sum\limits_{i \neq i'}^3 \left| C_{23}^{\nu_i \bar{\nu}_{i'} } \right|^2}{3 \left| C_{23}^{\rm SM} \right|^2}.
\end{align}
The related experimental data~\cite{Dattola:2021cmw,Belle:2017oht} and SM predictions~\cite{Blake:2016olu,Buras:2014fpa} provide $R^{\nu\bar{\nu}}_{\rm K}=2.4\pm 0.9$ for $B\rightarrow K^+ \nu\bar{\nu}$ and the upper limit $R^{\nu\bar{\nu}}_{\rm K^\ast}<2.7$ at the $90\%$ confidence level (CL) for $B\rightarrow K^\ast \nu\bar{\nu}$.

We collect the experimental results and SM predictions of $D^0 \to \ell^+ \ell^-$, $\tau \rightarrow \ell \rho^0$, $B \to \tau \nu$, $D_s \to \tau \nu$, and $\tau \to K \nu$, as well as the processes discussed above in table~\ref{tab:constraints1}. Following the same/analogical numerical calculations in the ordinary RPV-MSSM (see Refs.~\cite{Earl:2018snx,Hu:2020yvs}), the constraint from ${\cal B}(D^0 \to \mu^+ \mu^-)$ gives $|\lambda'_{223}|^2 < 0.31(m_{\tilde{b}_R}/1{\rm TeV})^2$, while the one from ${\cal B}(D^0 \to e^+ e^-)$ is negligible due to the small $m_e$, and the bound from ${\cal B}(\tau \rightarrow \ell \rho^0)$ provides $|\lambda'_{323}\lambda'^\ast_{2(1)23}| < 0.38(m_{\tilde{b}_R}/1{\rm TeV})^2$. 
Besides, $R_{133}$, $R_{223}$, and $R_{123}$, expressing the ratios of the measurement values to the SM predictions for
${\cal B}(B \to \tau \nu)$, ${\cal B}(D_s \to \tau \nu)$, and ${\cal B}(\tau \to K \nu)$ respectively, are also constrained. 
As for $\pi\to\ell\nu(\gamma)$ decay, similar to the formula in Ref.~\cite{Bryman:2021teu}, the bound (including loop corrections $h'$) is given by
\begin{align}
\frac{1+\eta_{\mu\mu}+h'_{\mu\mu}}{1+\eta_{ee}+h'_{ee}}=1.0010(9),
\end{align}
which can be translated into $|\eta_{ee}+h'_{ee}| \lesssim 0.0028$ within the $2\sigma$ level for negligible $\eta_{\mu\mu}$ and $h'_{\mu\mu}$. 
The ${\cal V}$ matrix is also bounded by $\tau(\mu)$ decaying to charged leptons and neutrinos at the tree level, while both couplings ${\cal V}$ and $\lambda'$ are generally constrained by these decays as well as the charged lepton flavor violating (cLFV) decays at one-loop level, which will be addressed in section~\ref{sec:loopconstraint}. 

\begin{table}[t]
\centering
\setlength\tabcolsep{8pt}
\renewcommand{\arraystretch}{1.3}
\begin{tabular}{|ccc|}
\hline
Observations & SM predictions  & Experimental data \\
\hline
  ${\cal B}(K^+ \to \pi^+ \nu \bar\nu)$  & $(9.24\pm0.83)\times 10^{-11}$~\cite{Aebischer:2018iyb}      & $(1.7\pm1.1)\times 10^{-10}$~\cite{ParticleDataGroup:2020ssz}   \\
  ${\cal B}(B^+\rightarrow K^+ \nu\bar{\nu})$ & $(4.6\pm 0.5)\times 10^{-6}$~\cite{Blake:2016olu} & $(1.1\pm 0.4)\times 10^{-5}$~\cite{Dattola:2021cmw} \\
  ${\cal B}(B\rightarrow K^{\ast 0} \nu\bar{\nu})$ & $(9.2\pm 1.0)\times 10^{-6}$~\cite{Buras:2014fpa} & $< 2.7\times 10^{-5}$~\cite{Belle:2017oht}\\
  ${\cal B}(D^0 \to \mu^+ \mu^-)$ & $\lesssim 6\times 10^{-11}$~\cite{LHCb:2013jyo} & $<6.2\times 10^{-9}$~\cite{LHCb:2013jyo}\\
  ${\cal B}(\tau \rightarrow \mu \rho^0)$ & - & $<1.2\times 10^{-8}$~\cite{ParticleDataGroup:2020ssz}\\
  ${\cal B}(B \to \tau \nu)$ & $(9.47\pm1.82) \times 10^{-5}$~\cite{Nandi:2016wlp} & $(1.09\pm0.24) \times 10^{-4}$~\cite{ParticleDataGroup:2020ssz} \\
  ${\cal B}(D_s \to \tau \nu)$ & $(5.40 \pm 0.30)\%$~\cite{Hu:2020yvs} & $(5.48 \pm 0.23)\%$~\cite{ParticleDataGroup:2020ssz}\\
  ${\cal B}(\tau \to K \nu)$ & $(7.15 \pm 0.026)\times 10^{-3}$~\cite{Hu:2018lmk} & 
  $(6.96 \pm 0.10)\times 10^{-3}$~\cite{ParticleDataGroup:2020ssz}\\
\hline
	\end{tabular}
	\caption{Current status of the related processes, which can be affected by RPV-MSSMIS at tree level. The experimental upper limits are given at the $90\%$ CL.}
	\label{tab:constraints1}
\end{table}     

\subsection{Loop-level processes}
\label{sec:loopconstraint}
Here we investigate the loop-level constraints and focus on cLFV processes at first. We stress that the non-$\lambda'$ NP contributions to these decays, $\tau \to \ell \gamma$, $\mu \to e \gamma$, $\tau \to \ell^{(')} \ell \ell$ ($\ell'\neq\ell$), and $\mu \to e e e$, can be eliminated for the particular structures of (s)neutrino mass matrices. That is, only chiral mixing but no flavor mixing exists for the sneutrino sector and the neutrino sector involving RH ones (see discussions in Ref.~\cite{Zheng:2021wnu} and section~\ref{sec:num}). Then, we focus on the $\lambda'$ contributions to the cLFV decays. 
The branching fraction of the $\tau\to\ell\gamma$ decay is given by~\cite{deGouvea:2000cf}
\begin{align}
{\cal B}(\tau\to\ell\gamma)=\frac{\tau_\tau\alpha m^5_\tau}{4}(|A^L_2|^2+|A^R_2|^2),
\end{align}
where the effective couplings $A^L_2=-\frac{\lambda'_{\ell j3}\lambda^{\prime\ast}_{3j3}}{64\pi^2 m^2_{\tilde{b}_R}}$ and $A^R_2=0$. The limit of $m^2_\ell/m^2_\tau \to 0$ is adopted here and also for other cLFV processes. 
It is worth noting that the cLFV muon decay $\mu\to e\gamma$ constrains the coupling product $|\lambda'_{1j3}\lambda'_{2j3}|$ with a TeV scale $m_{\tilde{b}_R}$ very strongly for the experimental upper limit ${\cal B}(\mu\to e\gamma)_{\rm exp}<4.2\times 10^{-13}$ at the $90\%$ CL~\cite{ParticleDataGroup:2020ssz}. That is, simultaneous non-negligible $\lambda'_{1j3}$ and $\lambda'_{2j3}$ are not favored, and hence it is also a reason that only non-negligible $\lambda'_{1j3}$ or $\lambda'_{2j3}$ are considered at a time in this work. Then $\mu \to e \gamma$, $\mu \to e e e$, and $\tau \to \ell^{'} \ell \ell$ will not be taken into consideration. 
The remained cLFV processes to be considered are $\tau\to\mu\gamma$, $\tau\to e\gamma$, $\tau\to\mu\mu\mu$, and $\tau\to eee$ decays (see the relevant formulas in Ref.~\cite{Hu:2020yvs}), with the experimental upper limits ${\cal B}(\tau\to\mu\gamma)_{\rm exp}<4.2\times10^{-8}$, ${\cal B}(\tau\to e\gamma)_{\rm exp}<3.3\times10^{-8}$, ${\cal B}(\tau\to\mu\mu\mu)_{\rm exp}<2.1\times 10^{-8}$ and ${\cal B}(\tau\to eee)_{\rm exp}<2.7\times 10^{-8}$ at $90\%$ CL, respectively~\cite{ParticleDataGroup:2020ssz}\footnote{In this work, in order to consider all the process constraints at the $2\sigma$ level, we get the experiment bound 
${\cal B}(\tau\to\mu\gamma)_{\rm exp}<5.1\times10^{-8}$, 
${\cal B}(\tau\to e\gamma)_{\rm exp}<4\times10^{-8}$, 
${\cal B}(\tau\to\mu\mu\mu)_{\rm exp}<2.6\times 10^{-8}$, 
${\cal B}(\tau\to eee)_{\rm exp}<3.3\times 10^{-8}$ as well as $R^{\nu\bar{\nu}}_{\rm K^\ast}<3.3$ under the assumption that the uncertainties follow the Gaussian distribution~\cite{Buttazzo:2017ixm}.}. 

Next, we investigate the $B_s-\bar{B}_s$ mixing, which is mastered by 
\begin{align}
{\cal L}_{\rm eff}^{b\bar{s}b\bar{s}}=(C_{B_s}^{\text{SM}}+C_{B_s}^{\text{NP}})
(\bar{s} \gamma_{\mu} P_L b)(\bar{s} \gamma^{\mu} P_L b)+{\rm h.c.},
\end{align}
where the SM contribution is
$C_{B_s}^{\rm SM} = -\frac{1}{4 \pi^2} G_F^2 m_W^2 \eta_t^2 S(x_t)$ with the defined function $S(x_t) \equiv \frac{x_t(4-11x_t+x_t^2)}{4(x_t-1)^2}+\frac{3x_t^3\log(x_t)}{2(x_t-1)^3}$, 
and the non-negligible NP contribution is
\begin{align}\label{eq:CBsMix}
C_{B_s}^{\rm NP} = \frac{1}{8i} \left(
\Lambda'^{\cal I}_{vv'}
D_2[m_{\tilde{\nu}^{\cal I}_v},m_{\tilde{\nu}^{\cal I}_{v'}},m_{b},m_{b}] 
+\Lambda'^{\cal N}_{vv'}
D_2[m_{\nu_v}, m_{\nu_{v'}},m_{\tilde{b}_R},m_{\tilde{b}_R}]
\right),
\end{align}
with $\Lambda'^{\cal I}_{vv'} \equiv \lambda'^{\cal I}_{v33} \lambda'^{\cal I \ast}_{v23} \lambda'^{\cal I}_{v'33} \lambda'^{\cal I \ast}_{v'23}$ and $\Lambda'^{\cal N}_{vv'} \equiv \lambda'^{\cal N}_{v33} \lambda'^{\cal N \ast}_{v23} \lambda'^{\cal N}_{v'33} \lambda'^{\cal N \ast}_{v'23}$. The  recently updated measurement by LHCb combined with previous ones induces $\Delta M_s^{\rm LHCb}=(17.7656 \pm 0.0057)$~${\rm ps}^{-1}$~\cite{LHCb:2021moh}\footnote{This combined result by LHCb is very close to the recent result averaged by HFLAV as $\Delta M_s^{\rm HFLAV-2021}=(17.765 \pm 0.006)$~${\rm ps}^{-1}$~\cite{HFLAV:2022pwe}, with the much improved precision compared to the previous average $\Delta M_s^{\rm HFLAV-2018}=(17.757 \pm 0.021)$~${\rm ps}^{-1}$~\cite{HFLAV:2019otj}.}, leading to the strong constraint along with the SM prediction $\Delta M_s^{\rm SM}=(18.4^{+0.7}_{-1.2})~{\rm ps}^{-1}$~\cite{DiLuzio:2019jyq}
\begin{align}\label{eq:DMsbound}
0.90 < |1+ C_{B_s}^{\rm NP}/C_{B_s}^{\rm SM}| < 1.11,
\end{align} at the $2\sigma$ level. 

Following the introduction of $B_s-\bar{B}_s$ mixing, we will mention the $B\to X_s \gamma$ decay, which is mastered by the electromagnetic dipole operator ${\cal O}_7=\frac{m_b}{e} (\bar{s} \sigma^{\mu\nu} P_{R} b) F_{\mu\nu}$. The $\lambda'$ contribution is given by 
\begin{align}\label{eq:C7rpv}
C_7^{\lambda'}=\frac{\sqrt{2} \lambda'^{\cal I}_{v3k}\lambda'^{\cal I \ast}_{v2k}}{144 G_F \eta_t m_{\tilde{\nu}^{\cal I}_v}^2}.
\end{align}
The non-$\lambda'$ ones can be referred to Ref.~\cite{deCarlos:1996yh} and are predicted to be negligible for the decoupled $\tilde{u}_i$. 
In order to fulfill the bound from the recent measured branching ratio ${\cal B}(B\rightarrow X_s \gamma)_{\rm exp}\times 10^4=3.43\pm 0.21\pm 0.07$~\cite{Amhis:2019ckw} and the SM prediction ${\cal B}(B\rightarrow X_s \gamma)_{\rm SM}\times 10^4=3.36\pm 0.23$~\cite{Misiak:2015xwa}, Eq.~\eqref{eq:C7rpv} implies a cancellation in $\lambda'^{\cal I}_{v33} \lambda'^{\cal I \ast}_{v23}$ for the (nearly) degenerate $m_{\tilde{\nu}}$. This cancellation is also beneficial for fulfilling the bounds of $B_s-\bar{B}_s$ mixing (see Eq.~\eqref{eq:CBsMix}).

Then we move on to the constraints from the purely leptonic decays of $Z$, $W$ bosons, and $\tau(\mu)$ leptons. The effective Lagrangian of $Z\to l^-_i l^+_j$ decay is given by~\cite{Arnan:2019olv}
\begin{align}\label{eq:Zll}
{\cal L}^{Zll}_{\rm eff}=\frac{e}{\cos\theta_W\sin\theta_W} \bar{l}_i \gamma^\mu \left(g_{l_L}^{ij} P_L + g_{l_R}^{ij} P_R \right) l_j Z_\mu,
\end{align}
where $g_{l_L}^{ij}=\delta^{ij}g_{l_L}^{\rm SM}+\delta g_{l_L}^{ij}$ and $g_{l_R}^{ij}=\delta^{ij}g_{l_R}^{\rm SM}+\delta g_{l_R}^{ij}$, with $g_{l_L}^{\rm SM}=-\frac{1}{2}+\sin^2\theta_W$ and $g_{l_R}^{\rm SM}=\sin^2\theta_W$. 
In the limit of $m_{l_i}/m_Z \to 0$, the corresponding branching fractions are
\begin{align} 
{\cal B}(Z\to l^-_i l^+_j)=\frac{m^3_Z}{6\pi v^2 \Gamma_Z} 
\left( |g_{l_L}^{ij}|^2+|g_{l_R}^{ij}|^2 \right) 
\end{align}
with $Z$ width $\Gamma_Z=2.495$~GeV~\cite{ParticleDataGroup:2020ssz}. For $i\neq j$, the branching ratio should be given by $\frac{1}{2}[{\cal B}(Z\to l^-_i l^+_j)+{\cal B}(Z\to l^-_j l^+_i)]$. 
The NP effective couplings contributed mainly by $\lambda'$ effects are expressed as $\delta g_{l_{L}}^{ij}=\frac{1}{32\pi^2} B^{ij}$ ($\delta g_{l_{R}}^{ij}=0$) here, and the formulas of $B^{ij}$ functions are collected in appendix~\ref{app:zdecays}. Then the measurements of the partial width ratios of $Z$ bosons, i.e., $\Gamma(Z\to\mu\mu)/\Gamma(Z\to ee)=1.0001(24)$, $\Gamma(Z\to\tau\tau)/\Gamma(Z\to\mu\mu)=1.0010(26)$, and $\Gamma(Z\to\tau\tau)/\Gamma(Z\to ee)=1.0020(32)$~\cite{ParticleDataGroup:2020ssz}, induce $|B^{11}|<0.36$ and $|B^{33}|<0.32$ when $\lambda'_{2jk}=0$ (Case A), and $|B^{22}|<0.35$ and $|B^{33}|<0.31$ when $\lambda'_{1jk}=0$ (Case B). Furthermore, the experimental upper limits, ${\cal}B(Z \to e\tau)<9.8\times10^{-6}$ and ${\cal}B(Z \to \mu\tau)<1.2\times10^{-5}$ at the $95\%$ CL~\cite{ParticleDataGroup:2020ssz}, make the bound $|B^{13}|^2+|B^{31}|^2<1.9^2$ and $|B^{23}|^2+|B^{32}|^2<2.1^2$ in Case A and B, respectively. 
We have checked that the additional effect on the $\theta_W$ extraction by $\eta$ and $h'$, which are in the range for $m^{\rm CDF-II}_W$ explanations, induces the extra influence on the bounds above up to $10^{-3}$, so this effect can be omitted safely. 

The constraints from the purely leptonic decays of $W$ boson  can be covered by the stronger ones from $\mu\to e\bar{\nu}_e\nu_\mu$ and $\tau\to\ell\bar{\nu}_{\ell}\nu_\tau$ decays. The fraction ratios of these lepton decays, i.e., ${\cal B}(\tau\to\mu\bar{\nu}_\mu\nu_\tau)/{\cal B}(\tau\to e\bar{\nu}_e\nu_\tau)$, ${\cal B}(\tau\to e\bar{\nu}_e\nu_\tau)/{\cal B}(\mu\to e\bar{\nu}_e\nu_\mu)$, and ${\cal B}(\tau\to\mu\bar{\nu}_\mu\nu_\tau)/{\cal B}(\mu\to e\bar{\nu}_e\nu_\mu)$, 
make the bounds~\cite{Bryman:2021teu} on the model parameters, which can be expressed as
\begin{align}\label{eq:cons_lepton}
\frac{1+\eta_{\mu\mu}+\delta^{\rm B X} h'_{\mu\mu}}
{1+\eta_{ee}+\delta^{\rm A X} h'_{ee}}&=1.0018(14),  \notag\\
\frac{1+\eta_{\tau\tau}+ h'_{\tau\tau}}
{1+\eta_{\mu\mu}+\delta^{\rm B X} h'_{\mu\mu}}&=1.0010(14),  \notag\\
\frac{1+\eta_{\tau\tau}+ h'_{\tau\tau}}
{1+\eta_{ee}+\delta^{\rm A X} h'_{ee}}&=1.0029(14),
\end{align} 
where X is A(B) for Case A(B). We only consider the $Wl\nu_l$-vertex, which has the interference with the SM contribution, but neglect the LFV-vertex  $Wl\nu_{l'}$ and $Zll'$, which can be embedded in $l\to l'\bar{\nu_i}\nu_j$ process. With the last two formulas of Eq.~\eqref{eq:cons_lepton} combined with $|\eta_{\mu\mu}(h'_{\mu\mu})|\lesssim 10^{-4}$, we should keep 
$|\eta_{\tau\tau}+ h'_{\tau\tau}|\lesssim 0.0018$ and 
$|\eta_{\tau\tau}+ h'_{\tau\tau}|\lesssim|\eta_{ee}+\delta^{\rm A X} h'_{ee}|$ at the $2\sigma$ level.

\section{Numerical analyses}
\label{sec:num}
In this section, we begin to study the numerical explanations for $B$-physics anomalies, $(g-2)_\mu$ and $m_W$ shift in RPV-MSSMIS. With the data of neutrino oscillation~\cite{Esteban:2020cvm}, we consider normal ordering and zero $\delta_{\rm CP}$. Then the three light neutrinos have masses $\{0,0.008,0.05\}$~eV with $m_{\nu_l} \approx \{0, \sqrt{\Delta m^2_{21}}, \sqrt{\Delta m^2_{31}} \}$~\cite{Alvarado:2020lcz}. The sets of fixed model parameters are collected in table~\ref{tab:input}.
\begin{table}[t]
\centering
\setlength\tabcolsep{8pt}
\renewcommand{\arraystretch}{1.3}
\begin{tabular}{|cc||cc|}
        \hline
		Parameters & Sets & Parameters  & Sets \\
		\hline
  $\tan\beta$  & $15$      & $Y_\nu$   & $\text{diag}(0.41,0.11,0.10)$\\
		$M_1$  & $320$~GeV & $M_R$     & $\text{diag}(1,1,1)$~TeV \\
		$M_2$  & $350$~GeV & $B_{M_R}$ & $\text{diag}(0.5,0.5,0.5)~\text{TeV}^2$    \\
		$\mu$  & $450$~GeV & $m_{\tilde{L}'_i}$   & $\text{diag}(360,350,350)$~GeV \\
    \hline
	\end{tabular}
	\caption{The sets of fixed model parameters, defined at $\mu_{\rm NP}$ scale.}
	\label{tab:input}
\end{table}
The diagonal inputs of $Y_\nu$, $M_R$, $m_{\tilde{L}'}$, and  $B_{M_R}$ induce no flavor mixings in sneutrino and neutrino sectors (when RH neutrinos engage), which are beneficial for fulfilling the bounds of cLFV decays (see appendix~\ref{app:nmat} for the particular discussions). Besides, the input values shown in table~\ref{tab:input} can induce a diagonal $\eta=-\text{diag}(2.53,0.18,0.15)\times 10^{-3}$. The values of $m_{\tilde{L}'_i}$ induce $m_{\tilde{\nu}_1(\tilde{l}_1)}=348(352)$~GeV, which are larger than the masses of the lightest neutralino and chargino, as $307$~GeV and $325$~GeV, respectively, and they are in accord with the constraints discussed in section~\ref{sec:cons_direct}. 
The remained parameters, $m_{\tilde{b}_R}$, $\lambda'_{323}$, $\lambda'_{333}$, $\lambda'_{1(2)23}$, and $\lambda'_{1(2)33}$ in Case A (B), can vary freely in the ranges considered. 

\begin{figure}[htbp]
	\centering
\includegraphics[width=0.95\textwidth]{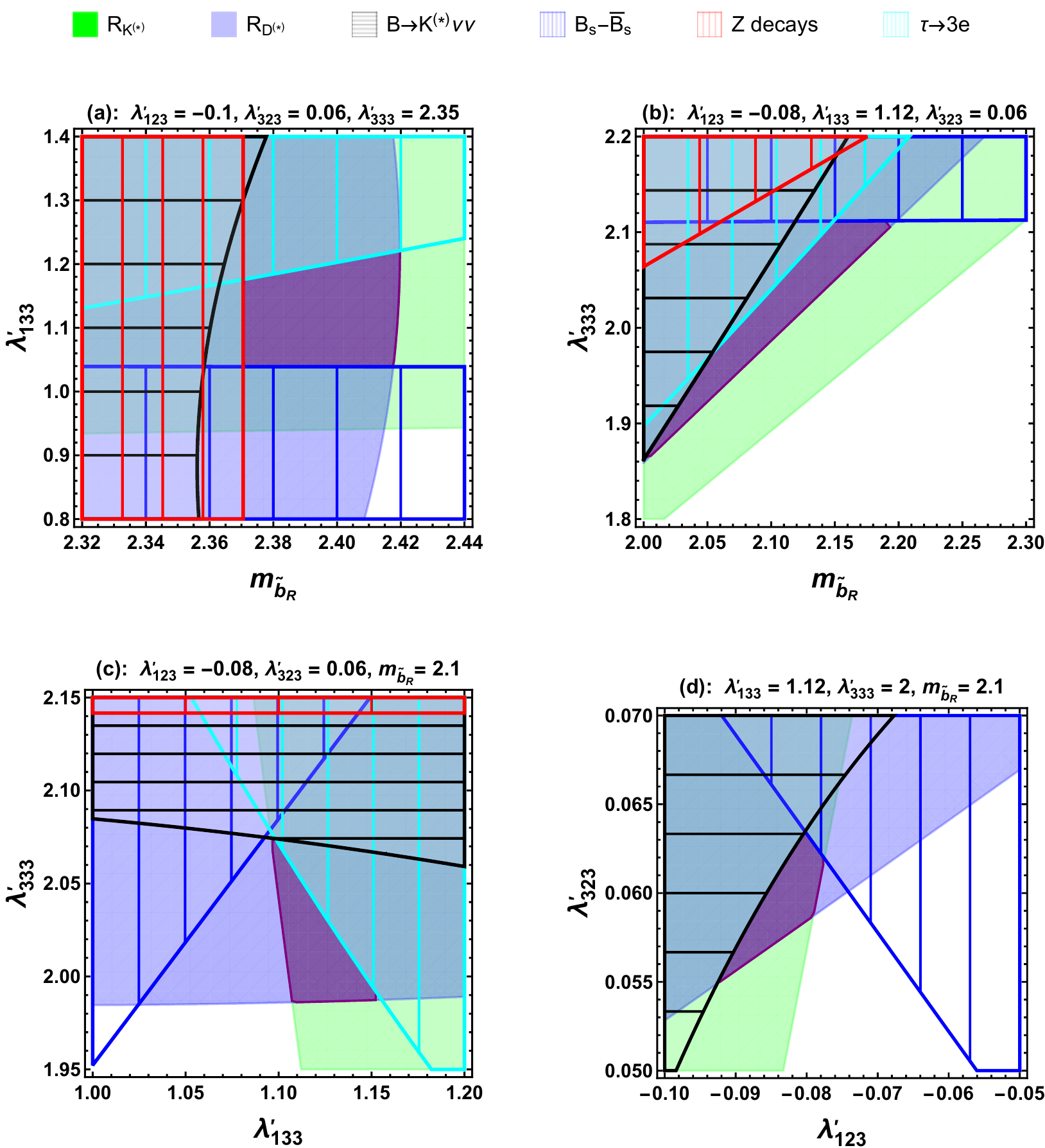}
	\caption{The experimental allowed regions for explaining the $R_{K^{(\ast)}}$ and $R_{D^{(\ast)}}$ anomalies in Case A. The masses $m_{\tilde{b}_R}$ are given in units of TeV. The $2\sigma$ favored areas for $R_{K^{(\ast)}}$ and $R_{D^{(\ast)}}$ measurements are denoted by green and blue, respectively. The hatched areas filled with the black-horizontal, red-vertical, cyan-vertical, and blue-vertical lines are exculded at the $2\sigma$ level by $B\rightarrow K^{(\ast)}\nu\bar{\nu}$, $Z\to l^-_i l^+_j$, $\tau \to e e e$ decays, and $B_s-\bar{B}_s$ mixing, respectively. The common areas are denoted by purple.}\label{fig:caseA}
\end{figure}

Next, with the input values given above and all $\lambda'$ couplings as real numbers, we can get some numerical results of Wilson coefficients and observables which show the possibilities for simultaneous explanations of the anomalies. They are given as follows,
\begin{align}\label{eq:number}
& C_{9e(\mu)}^{\rm NP}=-C_{10e(\mu)}^{\rm NP} \approx -1.604(-1.623) \lambda'_{1(2)23}\lambda'_{1(2)33},     \notag \\
& C_{B_s}^{\rm NP}/C_{B_s}^{\rm SM} \approx 76(\lambda'_{\ell 23}\lambda'_{\ell 33}+\lambda'_{323}\lambda'_{333})^2, \notag\\
& B^{33} \approx \frac{1{\rm ~TeV^2}}{m^2_{\tilde{b}_R}} \left[0.146 \log (m_{\tilde{b}_R}/1{\rm ~TeV})+0.2 \right]\lambda'^2_{333}, 
\end{align}
where only the dominant parts are kept. To be favored by  $R_{K^{(\ast)}}$ data at the $2\sigma$ level, there are $-0.355\lesssim\lambda'_{123}\lambda'_{133}\lesssim-0.106$ in Case A and $0.117\lesssim\lambda'_{223}\lambda'_{233}\lesssim0.314$ in Case B, through the first formula in Eq.~\eqref{eq:number}, which is induced by Eq.~\eqref{eq:Wcsbsll}. As for the second formula in Eq.~\eqref{eq:number}, valid for $1.5\leqslant m_{\tilde{b}_R}\leqslant 10$~TeV, we get $|\lambda'_{\ell 23}\lambda'_{\ell 33}+\lambda'_{323}\lambda'_{333}|\lesssim0.038$ with Eq.~\eqref{eq:DMsbound}. This result  implies that the $B_s-\bar{B}_s$ mixing bound demands the canceling between $\lambda'_{\ell 23}\lambda'_{\ell 33}$ and $\lambda'_{323}\lambda'_{333}$. In the last formula, the edge value of $\lambda'_{333}$ near the $Z\to\tau^-\tau^+$ bound can be gotten. With the rough NP features above, we will study them in detail.

In Case A, we can see that the $R_{K^{(\ast)}}$ and $R_{D^{(\ast)}}$ anomalies can be simultaneously explained at the $2\sigma$ level, although the overlaps are narrow, as shown in figure~\ref{fig:caseA}. The dominant constraints are from $B\to K^{(\ast)}\nu\bar{\nu}$, $Z$ leptonic decays, $\tau \to e e e$ decays, and $B_s-\bar{B}_s$ mixing. In figure~\ref{fig:caseA}(a), when $\lambda'$ couplings except $\lambda'_{133}$ are fixed as shown in the figure, the lower limit of $m_{\tilde{b}_R}$ is constrained by the $Z$ leptonic decays, i.e., $Z\to \tau^- \tau^+$ specifically, while $\tilde{b}_R$ should also be lighter than around $2.42$~TeV here to explain $R_{D^{(\ast)}}$ data. Accordingly, the favored region for $R_{K^{(\ast)}}$ explanations is broad and nearly independent of $m_{\tilde{b}_R}$, and it only demands $\lambda'_{133} \gtrsim 1$. In figure~\ref{fig:caseA}(b), we set $\lambda'_{133}=1.12$ and decrease $|\lambda'_{123}|$ slightly. The contributions to $R_{K^{(\ast)}}$ by $m_{\tilde{b}_R}$ cannot be omitted here because the chargino-sneutrino one is weakened slightly. The values of $(m_{\tilde{b}_R},\lambda'_{333})$ in the overlap get smaller, and then we get $-h'_{ee}=3.3\times 10^{-4}$ for $m_{\tilde{b}_R}=2.1$~TeV. The NP prediction $m_W^{\rm NP}$ without pure-MSSM effects is given by $m_W^{\rm NP} \approx m_W^{\rm SM} [1-0.20(\eta_{ee}+\eta_{\mu\mu}+h'_{ee})]$ (similar to the formula in Ref.~\cite{Fernandez-Martinez:2016lgt}), inducing the value $m_W^{\rm NP}=80.412$~GeV, which can explain the CDF-II measurement at the $2\sigma$ level. Keeping the values of $m_{\tilde{b}_R}=2.1$~TeV, relevant overlaps are also found in figure~\ref{fig:caseA}(c) and figure~\ref{fig:caseA}(d).

\begin{figure}[htbp]
	\centering
\includegraphics[width=0.95\textwidth]{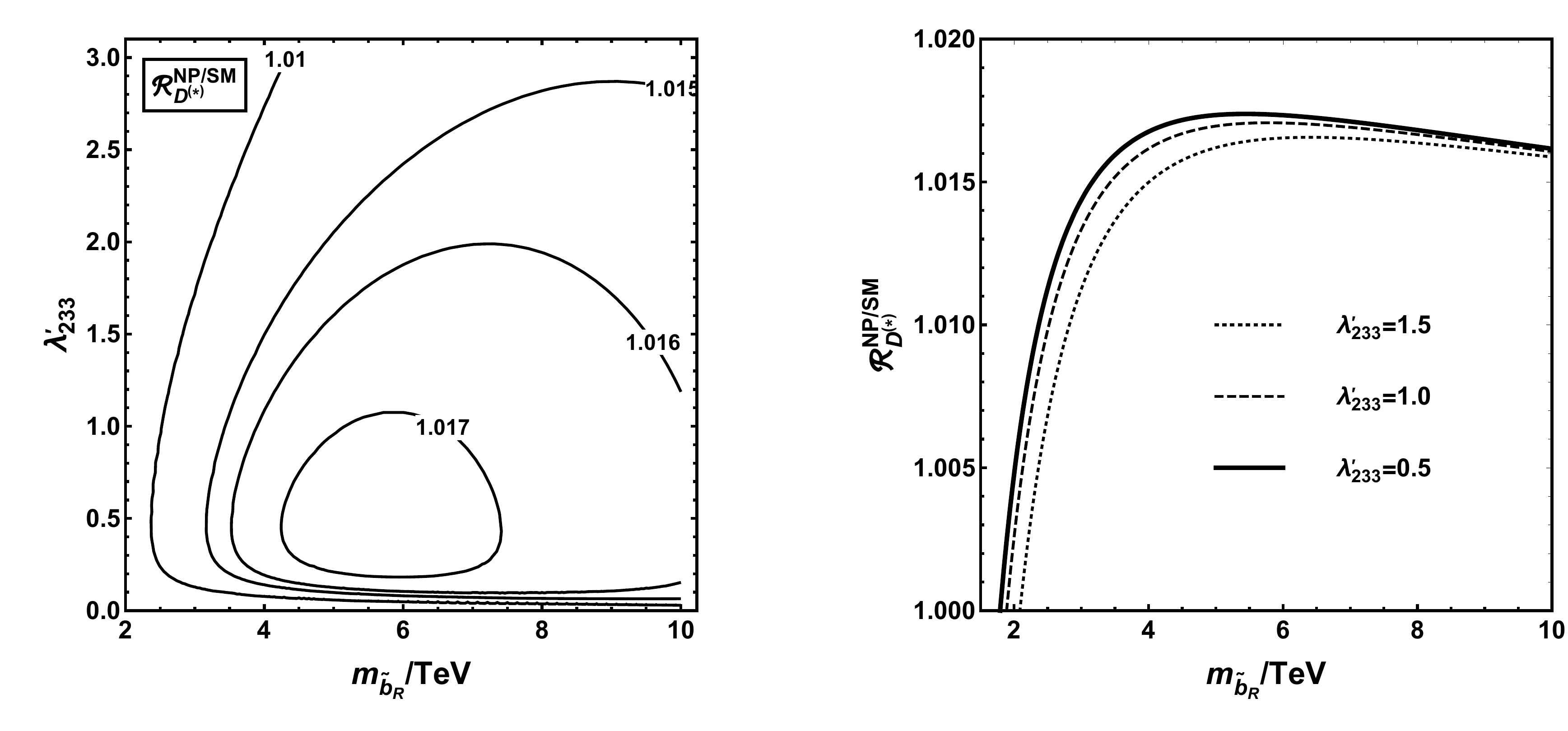}
	\caption{The ratio ${\cal R}^{\rm NP/SM}_{D^{(\ast)}}$ varies with $m_{\tilde{b}_R}$ and $\lambda'_{233}$ in Case B. In the right panel, $\lambda'_{233}$ is set as $1.5$ (dotted), $1$ (dashed), and $0.5$ (solid), respectively.}\label{fig:caseB}
\end{figure}

Then we move on to Case B. In this case, we will show that, given the stringent bounds from the $B_s-\bar{B}_s$ mixing as well as the constraints of $Z\to \tau^- \tau^+$ decay, the anomalies of  $R_{K^{(\ast)}}$ and $R_{D^{(\ast)}}$ cannot be explained simultaneously at the $2\sigma$ level. From Eq.~\eqref{eq:RDratio2}, one can see that the $R_{D^{(\ast)}}$ data explanation favors large $|\lambda'_{333}|$, positive $\lambda'_{323}\lambda'_{333}$, and negative $\lambda'_{\ell 23}\lambda'_{\ell 33}$. However, positive $\lambda'_{223}\lambda'_{233}$ is needed for explaining $R_{K^{(\ast)}}$. Thus, with Eq.~\eqref{eq:number}, we let $\lambda'_{223}$ be the edge value $0.117/\lambda'_{233}$ for $R_{K^{(\ast)}}$ explanations and $\lambda'_{323}=(-0.117+0.038)/\lambda'_{333}$ for the bound edge of $B_s-\bar{B}_s$ mixing, using $\lambda'_{333}=\sqrt{0.31}(m_{\tilde{b}_R}/1{\rm ~TeV})/\left[0.146 \log (m_{\tilde{b}_R}/1{\rm ~TeV})+0.2 \right]^{\frac{1}{2}}$ as the edge value near the $Z\to\tau^-\tau^+$ bound. Then the prediction of 
${\cal R}^{\rm NP/SM}_{D^{(\ast)}}$ can be given as the function of $(m_{\tilde{b}_R},\lambda'_{233})$, as shown in figure~\ref{fig:caseB}. 
For $0.5 \lesssim \lambda'_{233} \lesssim 1$, we find that the ratio increases with $m_{\tilde{b}_R}$ increasing up to  around $6$~TeV because the edge value of $\lambda'_{333}$ constrained by $Z$ decays gets relatively large for heavier $m_{\tilde{b}_R}$. Then the ratio decreases for $m_{\tilde{b}_R}\gtrsim 6$~TeV. From the right panel of figure~\ref{fig:caseB}, it is obvious that ${\cal R}^{\rm NP/SM}_{D^{(\ast)}}$ cannot be raised higher than $1.02$, with below the lower limit $1.05$ of $2\sigma$ fit. Thus, the simultaneous $2\sigma$-level explanation of $R_{K^{(\ast)}}$ and $R_{D^{(\ast)}}$ anomalies is impossible in Case B. While we can still explain $R_{K^{(\ast)}}$ and other anomalies in the $b\to s\mu^+\mu^-$ process, which will be shown in table~\ref{tab:benchmark}, with the benchmark point in Case A collected as well.

In table~\ref{tab:benchmark}, as mentioned before, for the point in Case A, both $R_{K^{(\ast)}}$ and $R_{D^{(\ast)}}$ are in the $2\sigma$ ranges of the experimental measurements. In Case B(I), we utilize the relevant point to fulfill  $R_{K^{(\ast)}}$ data at $2\sigma$, and raise the ratio ${\cal R}^{\rm NP/SM}_{D^{(\ast)}}$. However, this increase cannot reach the $2\sigma$ accordance region as we predicted. The point in Case B(II) can explain the $b\to s \ell^+\ell^-$ anomalies at the $2\sigma$ level without considering  $R_{D^{(\ast)}}$ data. Besides, in all the cases, $(g-2)_\mu$ data and CDF-II $m_W$ shift favor the points, which are also  allowed by relevant constraints shown in section~\ref{sec:constraints}. Due to the contributions of $\lambda'$ diagrams, there are slight differences among the predicted $\delta a^{\rm NP}_{\mu}$ in the three cases. In case the CDF-II result is not supported by future measurements, we can adjust the coupling $Y_\nu$ in table~\ref{tab:input} into, e.g., $\text{diag}(0.12,0.11,0.10)$, which induces $\eta_{ee}=-2.2\times10^{-4}$ to fulfill the current global  $m_W=80.379(12)$~GeV~\cite{ParticleDataGroup:2020ssz} and relevant measurements of $\sin \theta_W$ mentioned in section~\ref{sec:mW}. We have checked that the explanations for the other anomalies in Case A, Case B(I), and Case B(II) are still applicable. 

Before the end of this section, we discuss probes for RPV-MSSMIS in future experiments. 
The observation of LFV in $Z$ decays would provide  indisputable evidence for NP, and especially, $Z\to e\tau$ and $Z\to \mu\tau$ decays can connect to the $\tau$ flavor, which can probe the leptoquark or RPV-SUSY models. With a TeV scale of $m_{\tilde{b}_R}$ and relevant $\lambda'_{i33}$ values in table~\ref{tab:benchmark}, the branching ratios ${\cal B}(Z\to e\tau)$ and ${\cal B}(Z\to \mu\tau)$ are predicted as ${\cal O}(10^{-7})$ scale, which can reach the sensitivities ${\cal O}(10^{-9})$ at FCC-ee~\cite{FCC:2018byv}.
Besides, the model predicts the branching ratios of the cLFV processes, $\tau \to \ell \gamma$ and $\tau \to \ell\ell\ell$ with the scale of ${\cal O}(10^{-9}-10^{-8})$, and they can reach the future sensitivities by Belle II ($50$~${\rm ab}^{-1}$)~\cite{Belle-II:2018jsg} and FCC-ee~\cite{FCC:2018byv}. Even in $\tilde{b}_R$-decoupled scenario, which is only favored by $b\to s\ell^+\ell^-$ and muon $g-2$ explanations, the heavy neutrinos with TeV masses can also reach future collider searches~\cite{Abdullahi:2022jlv,Arguelles:2022xxa}. For instance, in the multi-lepton channel at future colliders ($\sqrt{s}=27$~TeV), active-sterile mixing as small as $|{\cal V}_{lN}|^2\sim {\cal O}(10^{-3}-10^{-2})$ can be probed at the $95\%$ CL for heavy neutrino masses in the range $700<m_{\nu_N}<3500$~GeV with $15$~${\rm ab}^{-1}$ of data~\cite{Pascoli:2018heg}. 

\begin{table}[htbp]
\centering
\setlength\tabcolsep{8pt}
\renewcommand{\arraystretch}{1.3}
\begin{tabular}{|c|ccc|}
        \hline
		~ & Case A & Case B(I) & Case B(II)    \\
\hline
$\lambda'_{\ell 23}$ & $-0.08$ & $0.1$    & $0.1$  \\
$\lambda'_{\ell 33}$ & $1.12$  & $1.35$    & $2$   \\
$\lambda'_{3 23}$    & $0.06$  & $-0.04$  &$-0.1$  \\
$\lambda'_{3 33}$    & $2$     & $2.6$    & $2$    \\
$m_{\tilde{b}_R}/{\rm TeV}$& $2.1$  & $3$    & $3$ \\
$C_{9\ell}^{\rm NP}=-C_{10\ell}^{\rm NP}$ & $0.172$ & $-0.196$ & $-0.295$ \\
$C_{9{\rm U}}^{\rm NP}$ & $-0.157$ & $-0.158$ &$-0.001$ \\
${\cal R}^{\rm NP/SM}_{D^{(\ast)}}$ & $1.055$ & $1.006$ & $0.983$ \\	
$m_W^{\rm NP}/{\rm GeV}$ & $80.412$ & $80.411$  &$80.416$ \\
$\delta a^{\rm NP}_{\mu}\times 10^9$ & $1.361$ & $1.369$ & $1.377$\\
\hline
	\end{tabular}
	\caption{The benchmark points in Case A ($\ell=1$ and $e$) and B ($\ell=2$ and $\mu$). The Wilson coefficient $C_{9{\rm U}}^{\rm NP}$ provides the lepton flavor universal contribution (see Ref.~\cite{Zheng:2021wnu}).}
	\label{tab:benchmark}
\end{table}

\section{Conclusions}
\label{sec:conclusion}
The recently reported anomalies in $R_{D^{(\ast)}}$, $R_{K^{(\ast)}}$ with other $b\to s\ell^+\ell^-$ observables, e.g., $P'_5$, ${\cal B}(B_s\to \phi\mu^+\mu^-)$, as well as the enduring muon $g-2$, have shown that LFUV effects beyond the SM may exist. While the more recent precision measurement of $W$ mass, if confirmed by other experiments, will profoundly change the situation of NP search. Interestingly, the LFUV NP can affect $m_W$ prediction through muon decays, so it is straightforward to make a simultaneous investigation on the $B$-physics anomalies, $(g-2)_\mu$, and $m_W$ in NP models.

In this work, we study these anomalies mentioned above in RPV-MSSMIS, which is the framework providing $\lambda'\hat L \hat Q \hat D$ interaction involved with the (s)neutrino chiral mixing for explaining $B$-physics anomalies and increasing $m_W$ prediction, which is also contributed by the non-unitarity $\eta_{ee}$ in the seesaw sector. We consider nonzero $\lambda'_{1(2)jk}$ at a time in Case A(B), and find that the deviations of $R_{K^{(\ast)}}$, $R_{D^{(\ast)}}$, and $m_W$ from SM predictions can be reduced to the $2\sigma$ level simultaneously in Case A. In Case B, the combined $2\sigma$-level explanation of $b\to s\ell^+\ell^-$ anomaly and $m_W$-shift can be achieved, while NP effects on $R_{D^{(\ast)}}$ are weak. Moreover, $(g-2)_\mu$ data can be explained in both cases, which also fulfill neutrino oscillation data, the relevant constraints at collider, and a series of flavor-physics bounds from $B\to K^{(\ast)}\nu\bar{\nu}$, $Z$ leptonic decays, cLFV decays, $B_s-\bar{B}_s$ mixing, etc. Moreover, the LFV effects on the $Z$-boson and $\tau$ decays as well as the TeV scale heavy neutrinos in this model, can be testable in future experiments.     

\section*{Acknowledgements}
We thank Seishi Enomoto for valuable discussions.
This work is supported in part by the National Natural Science Foundation of China under Grant No. 11875327, the Fundamental Research Funds for the Central Universities, and the Sun Yat-Sen University Science Foundation. F.C. is also supported by the CCNU-QLPL Innovation Fund (QLPL2021P01).

\appendix

\section{One-loop box contributions to $b\to s\ell^+\ell^-$ in RPV-MSSMIS}
\label{app:bsllCbox}
In this section, the whole Wilson coefficients from the one-loop $b\to s\ell^+\ell^-$ boxes in RPV-MSSMIS are listed.

The LH-quark-current contributions from chargino boxes to $b\to s\ell^+\ell^-$ process are given by
\begin{align}\label{eq:chaWcs}
C^{\chi^\pm}_{9\ell}=&-C^{\chi^\pm}_{10\ell}=-\frac{\sqrt{2}\pi^2 i}{2G_F \eta_t e^2} \Bigl(
g^2_2 K_{i3} K^{\ast}_{i2} V^{\ast}_{m1}V_{n1}
(g_2V_{m1}\tilde{\cal V}^{\cal I}_{v\ell}-
V_{m2}Y^{\cal I}_{\ell v}) \notag\\
&(g_2V^{\ast}_{n1}\tilde{\cal V}^{\cal I}_{v\ell}-
V^{\ast}_{n2}Y^{\cal I}_{\ell v}) 
D_2[m_{\tilde{\nu}^{\cal I}_{v}},m_{\chi^\pm_m},m_{\chi^\pm_n},m_{\tilde{u}_{Li}}] \notag\\
&+y^2_{u_i} K_{i3}K^{\ast}_{i2} V^{\ast}_{m2}V_{n2}
(g_2V_{m1}\tilde{\cal V}^{\cal I}_{v\ell}-
V_{m2}Y^{\cal I}_{\ell v}) \notag\\
&(g_2V^{\ast}_{n1}\tilde{\cal V}^{\cal I}_{v\ell}-
V^{\ast}_{n2}Y^{\cal I}_{\ell v}) 
D_2[m_{\tilde{\nu}^{\cal I}_{v}},m_{\chi^\pm_m},m_{\chi^\pm_n},m_{\tilde{u}_{Ri}}] \notag\\
&-\lambda'^{\cal I}_{v3k} \lambda'^{\cal I \ast}_{v'2k} (g_2 V^{\ast}_{m1} \tilde{\cal V}^{\cal I}_{v\ell}-V^{\ast}_{m2} Y^{\cal I}_{\ell v})
(g_2 V_{m1} \tilde{\cal V}^{\cal I}_{v'\ell}-V_{m2} Y^{\cal I}_{\ell v'}) D_2[m_{\tilde{\nu}^{\cal I}_{v}},m_{\tilde{\nu}^{\cal I}_{v'}},m_{\chi^\pm_m},m_{d_k}] \notag\\
&-\tilde{\lambda}'_{\ell ik} \tilde{\lambda}'^{\ast}_{\ell jk} g^2_2 K_{i3}K^{\ast}_{j2} |V_{m1}|^2 D_2[m_{\tilde{u}_{Li}},m_{\tilde{u}_{Lj}},m_{\chi^\pm_m},m_{d_k}] \notag\\
&+\tilde{\lambda}'_{\ell ik} \lambda'^{\cal I \ast}_{v2k} (g_2 K_{i3} V^{\ast}_{m1})(g_2 V_{m1} \tilde{\cal V}^{\cal I}_{v\ell}-V_{m2} Y^{\cal I}_{\ell v}) D_2[m_{\tilde{\nu}^{\cal I}_{v}},m_{\tilde{u}_{Li}},m_{\chi^\pm_m},m_{d_k}] \notag\\
&+\tilde{\lambda}'^{\ast}_{\ell ik} \lambda'^{\cal I}_{v3k} (g_2 K^{\ast}_{i2} V_{m1})(g_2 V^{\ast}_{m1} \tilde{\cal V}^{\cal I}_{v\ell}-V^{\ast}_{m2} Y^{\cal I}_{\ell v}) D_2[m_{\tilde{\nu}^{\cal I}_{v}},m_{\tilde{u}_{Li}},m_{\chi^\pm_m},m_{d_k}]
\Bigr),
\end{align}
where the Yukawa couplings $y_{u_i}={\sqrt{2}m_{u_i}}/{v_u}$ and $Y^{\cal I}_{\ell v}\equiv{(Y_\nu)}_{j\ell} \tilde{\cal V}^{\cal I \ast}_{v(j+3)}$. While the corresponding RH-quark-current contributions are
\begin{align}\label{eq:chaWcsRH}
C^{\prime\chi^\pm}_{9\ell}=&-C^{\prime\chi^\pm}_{10\ell}=-\frac{\sqrt{2}\pi^2 i}{2G_F \eta_t e^2}
\lambda'^{\cal I}_{vi2} \lambda'^{\cal I \ast}_{v'i3} (g_2 V^{\ast}_{m1} \tilde{\cal V}^{\cal I}_{v\ell}-V^{\ast}_{m2} Y^{\cal I}_{\ell v}) \notag\\
&(g_2 V_{m1} \tilde{\cal V}^{\cal I}_{v'\ell}-V_{m2} Y^{\cal I}_{\ell v'}) D_2[m_{\tilde{\nu}^{\cal I}_{v}},m_{\tilde{\nu}^{\cal I}_{v'}},m_{\chi^\pm_m},m_{d_i}].
\end{align}

The contributions of $W/H^\pm$ (means $W$ with $W$ Goldstones or charged Higgs box involved) box diagrams to $b\to s\ell^+\ell^-$ process are given by
\begin{align}\label{eq:WHWcs}
C^{W/H^\pm}_{9\ell}=&-C^{W/H^\pm}_{10\ell}=-\frac{\sqrt{2}\pi^2 i}{2G_F \eta_t e^2} \Bigl(
y^2_{u_i} K_{i3}K^{\ast}_{i2} Z^2_{H_{h2}}Z^2_{H_{h'2}} |Y^{\cal N}_{\ell v}|^2 D_2[m_{\nu_v},m_{u_i},m_{H_h},m_{H_{h'}}] \notag\\
&-4g^2_2 m_{u_i}y_{u_i} m_{\nu_v} K_{i3}K^{\ast}_{i2} Z^2_{H_{h2}} {\text{Re}}({\cal V}^{\ast}_{v\ell} Y^{\cal N \ast}_{\ell v}) D_0[m_{\nu_v},m_{u_i},m_W,m_{H_{h}}] \notag\\
&+5g^4_2 K_{i3}K^{\ast}_{i2} |{\cal V}^{\ast}_{v\ell}|^2 D_2[m_{\nu_v},m_{u_i},m_W,m_W] \notag\\
&+Z^2_{H_{h2}} Y^{\cal N \ast}_{\ell v} Y^{\cal N}_{\ell v'} \lambda'^{\cal N}_{v3k} \lambda'^{\cal N \ast}_{v'2k} D_2[m_{\nu_v},m_{\nu_{v'}},m_{H_h},m_{\tilde{d}_{Rk}}] \notag\\
&-2g^2_2 m_{\nu_v} m_{\nu_{v'}} {\cal V}_{v\ell} {\cal V}^{\ast}_{v'\ell} \lambda'^{\cal N}_{v3k} \lambda'^{\cal N \ast}_{v'2k} D_0[m_{\nu_v},m_{\nu_{v'}},m_W,m_{\tilde{d}_{Rk}}] \notag\\
&+2 m_{\nu_v}m_{\nu_{v'}} Z^2_{H_{h2}} Y^{\cal N}_{\ell v} Y^{\cal N \ast}_{\ell v'} 
\lambda'^{\cal N}_{v3k} \lambda'^{\cal N \ast}_{v'2k}
D_0[m_{\nu_v},m_{\nu_{v'}},m_{H_h},m_{\tilde{d}_{Rk}}] \notag\\
&+2 m_{u_i}m_{u_j} y_{u_i}y_{u_j} K_{i3}K^{\ast}_{j2} Z^2_{H_{h2}}
\tilde{\lambda}'_{\ell ik} \tilde{\lambda}'^{\ast}_{\ell jk}
D_0[m_{u_i},m_{u_j},m_{H_h},m_{\tilde{d}_{Rk}}] \notag\\
&-g^2_2 {\cal V}^{\ast}_{v\ell} {\cal V}_{v'\ell} \lambda'^{\cal N}_{v3k} \lambda'^{\cal N \ast}_{v'2k}
D_2[m_{\nu_v},m_{\nu_{v'}},m_{W},m_{\tilde{d}_{Rk}}] \notag\\
&-g^2_2 K_{i3}K^{\ast}_{j2} \tilde{\lambda}'_{\ell ik} \tilde{\lambda}'^{\ast}_{\ell jk}
D_2[m_{u_i},m_{u_j},m_{W},m_{\tilde{d}_{Rk}}] \notag\\
&-2 m_{u_i}y_{u_i} m_{\nu_v} K_{i3} Z^2_{H_{h2}} Y^{\cal N \ast}_{\ell v} \tilde{\lambda}'_{\ell ik} 
\lambda'^{\cal N \ast}_{v2k}
D_0[m_{u_i},m_{\nu_v},m_{H_h},m_{\tilde{d}_{Rk}}] \notag\\
&-2 m_{u_i}y_{u_i} m_{\nu_v} K^{\ast}_{i2} Z^2_{H_{h2}} Y^{\cal N}_{\ell v} \tilde{\lambda}'^{\ast}_{\ell ik} 
\lambda'^{\cal N}_{v3k}
D_0[m_{u_i},m_{\nu_v},m_{H_h},m_{\tilde{d}_{Rk}}] \notag\\ 
&+g^2_2 K^{\ast}_{i2} {\cal V}^{\ast}_{v\ell} \tilde{\lambda}'^{\ast}_{\ell ik} 
\lambda'^{\cal N}_{v3k}
D_2[m_{u_i},m_{\nu_v},m_W,m_{\tilde{d}_{Rk}}] \notag\\
&+g^2_2 K_{i3} {\cal V}_{v\ell} \tilde{\lambda}'_{\ell ik} 
\lambda'^{\cal N \ast}_{v2k}
D_2[m_{u_i},m_{\nu_v},m_W,m_{\tilde{d}_{Rk}}]
\Bigr), \\
\ \notag\\
C^{\prime W/H^\pm}_{9\ell}=&-C^{\prime W/H^\pm}_{10\ell}=-\frac{\sqrt{2}\pi^2 i}{2G_F \eta_t e^2} \Bigl( 
-2 Z^2_{H_{h2}} Y^{\cal N\ast}_{\ell v} Y^{\cal N}_{\ell v'}
\lambda'^{\cal N}_{v'i2} \lambda'^{\cal N \ast}_{vi3} 
m_{\nu_v} m_{\nu_{v'}} \notag\\
&\times D_0[m_{\nu_v},m_{\nu_{v'}},m_{H_h},m_{\tilde{d}_{Li}}] 
- Z^2_{H_{h2}} Y^{\cal N}_{\ell v} Y^{\cal N\ast}_{\ell v'}
\lambda'^{\cal N}_{v'i2} \lambda'^{\cal N \ast}_{vi3}
D_2[m_{\nu_v},m_{\nu_{v'}},m_{H_h},m_{\tilde{d}_{Li}}] \notag\\
&+g^2_2 {\cal V}^{\ast}_{v\ell} {\cal V}_{v'\ell} 
\lambda'^{\cal N}_{vi2} \lambda'^{\cal N \ast}_{v'i3}
D_2[m_{\nu_v},m_{\nu_{v'}},m_W,m_{\tilde{d}_{Li}}] \notag\\
&+2 g^2_2 {\cal V}^{\ast}_{v\ell} {\cal V}_{v'\ell} 
\lambda'^{\cal N}_{v'i2} \lambda'^{\cal N \ast}_{vi3}
m_{\nu_v} m_{\nu_{v'}} 
D_0[m_{\nu_v},m_{\nu_{v'}},m_W,m_{\tilde{d}_{Li}}]
\Bigr),
\end{align}
where the mixing matrix elements $Z_{H_{12}}=-\sin \beta$, $Z_{H_{22}}=-\cos\beta$ with Goldstone mass $m_{H_1}=m_W$ and changed Higgs mass $m_{H_2}=m_{H^\pm}$, and $Y^{\cal N}_{\ell v}\equiv{(Y_\nu)}_{j\ell} {\cal V}_{v(j+3)}$.

The contributions of $4\lambda'$ box diagrams to $b\to s\ell^+\ell^-$ process are given by
\begin{align}
C^{4\lambda'}_{9\ell}=&-C^{4\lambda'}_{10\ell}=-\frac{\sqrt{2}\pi^2 i}{2G_F \eta_t e^2} \Bigl(
\tilde{\lambda}'_{\ell ik}  \tilde{\lambda}'^{\ast}_{\ell ik'}
\lambda'^{\cal N}_{v3k'} 
\lambda'^{\cal N \ast}_{v2k}
D_2[m_{\nu_v},m_{u_i},m_{\tilde{d}_{Rk}},m_{\tilde{d}_{Rk'}}] \notag\\
&+\tilde{\lambda}'_{\ell ik'}  \tilde{\lambda}'^{\ast}_{\ell ik}
\lambda'^{\cal I}_{v3k} 
\lambda'^{\cal I \ast}_{v2k'}
D_2[m_{\tilde{\nu}^{\cal I}_v},m_{\tilde{u}_{Li}},m_{d_k},m_{d_{k'}}]
\Bigr),\\
\ \notag\\
C^{\prime4\lambda'}_{9\ell}=&-C^{\prime4\lambda'}_{10\ell}=-\frac{\sqrt{2}\pi^2 i}{2G_F \eta_t e^2} \Bigl(
\tilde{\lambda}'_{\ell ij}  \tilde{\lambda}'^{\ast}_{\ell ij'}
\lambda'^{\cal N}_{vj2} 
\lambda'^{\cal N \ast}_{vj'3}
D_2[m_{\nu_v},m_{u_i},m_{\tilde{d}_{Lj}},m_{\tilde{d}_{Lj'}}] \notag\\
&-\tilde{\lambda}'_{\ell j'k} \tilde{\lambda}'^{\ast}_{\ell jk}
\tilde{\lambda}'_{ij2} \tilde{\lambda}'^{\ast}_{ij'3}
\bigl(D_2[m_{l_i},m_{\tilde{u}_{Lj}},m_{\tilde{u}_{Lj'}},m_{d_k}]
+D_2[m_{\tilde{l}_{Li}},m_{u_j},m_{u_{j'}},m_{\tilde{d}_{Rk}}] \bigr) \Bigr).
\end{align}

The contributions of neutralino box diagrams only contain RH-quark-current parts, which are given by
\begin{align}\label{eq:neuWcs}
C^{\prime \chi^0}_{9\ell}=&- C^{\prime \chi^0}_{10\ell}=-\frac{\sqrt{2}\pi^2 i}{2G_F \eta_t e^2} \Bigl(
\frac{1}{2} (g_1 {N_{n1}} + g_2 {N_{n2}})^2
\tilde{\lambda}'_{\ell i2} \tilde{\lambda}'^{\ast}_{\ell i3}
D_2[m_{\tilde{l}_{L\ell}},m_{\tilde{l}_{L\ell}},m_{u_i},m_{\chi^0_n}] \notag\\
&+\frac{2}{9} g^2_1 |{N_{n1}}|^2 \tilde{\lambda}'_{\ell i2} \tilde{\lambda}'^{\ast}_{\ell i3}
D_2[m_{u_i},m_{\chi^0_n},m_{\tilde{s}_{R}},m_{\tilde{b}_{R}}] \notag\\
&-\frac{1}{3} {N_{n1}} g_1 (g_1 {N_{n1}} + g_2 {N_{n2}})
\tilde{\lambda}'_{\ell i2} \tilde{\lambda}'^{\ast}_{\ell i3} 
 \bigl(D_2[m_{\tilde{l}_{L\ell}},m_{u_i},m_{\chi^0_n},m_{\tilde{s}_{R}}]+
D_2[m_{\tilde{l}_{L\ell}},m_{u_i},m_{\chi^0_n},m_{\tilde{b}_{R}}] \bigr)
\Bigr),
\end{align}
where $m_{\chi^0_n}$ is the neutralino mass after the diagonalization $N {\cal M}_{\chi^0} N^T={\cal M}_{\chi^0}^{\text{diag}}$.

In the formulas above, the Passarino-Veltman functions~\cite{Passarino:1978jh} $D_0$ and $D_2$ are defined as
\begin{align}
D_0[m_1,m_2,m_3,m_4] 
\equiv& \int \frac{d^4 k}{(2\pi)^4}\frac{1}{(k^2-m_1^2)(k^2-m_2^2)(k^2-m_3^2)(k^2-m_4^2)}, \notag \\
D_2[m_1,m_2,m_3,m_4] 
\equiv& \int \frac{d^4 k}{(2\pi)^4}\frac{k^2}{(k^2-m_1^2)(k^2-m_2^2)(k^2-m_3^2)(k^2-m_4^2)}. 
\end{align}

\section{The coupling functions in $Z\to l^{-}_i l^{+}_j$ process}
\label{app:zdecays}

With the effective Lagrangian of $Z\to l^-_i l^+_j$ process in Eq.~\eqref{eq:Zll}, the functions $B^{ij} \equiv (32 \pi^2) \delta g_{l_{L}}^{ij} $ are given by the following two parts~\cite{Earl:2018snx},
\begin{align}
B^{ij}_1 =& 3 \tilde{\lambda}'_{j33} \tilde{\lambda}'^{\ast}_{i33} \biggl\{-x_{\tilde{b}_R} (1 + \log x_{\tilde{b}_R} ) 
+ \frac{m_Z^2}{18 m^2_{\tilde{b}_R}} \biggl[(11 - 10 \sin^2\theta_W) \notag \\ 
&+ (6 - 8 \sin^2\theta_W)\log x_{\tilde{b}_R} + \frac{1}{10}(-9 + 16 \sin^2\theta_W)\frac{m_Z^2}{m_t^2} \biggr] \biggl\},   \notag\\
B^{ij}_2 =& \sum_{\ell=1}^2 \tilde{\lambda}'_{j\ell 3} \tilde{\lambda}'^{\ast}_{i\ell 3} \frac{m_Z^2}{m^2_{\tilde{b}_R}} \biggl[(1 - \frac{4}{3} \sin^2 \theta_W)(\log \frac{m_Z^2}{m^2_{\tilde{b}_R}} - i\pi -\frac{1}{3} )+
\frac{\sin^2 \theta_W}{9} \biggr].
\end{align}
Then there are $B^{ij}=B^{ij}_1+B^{ij}_2$. $B^{ij}_1$ is the dominant part due to the involved top quark.

\section{The numerical form of the (s)neutrino mixing matrix}
\label{app:nmat}

With the input set in table~\ref{tab:input}, the numerical form of the neutrino mixing matrix is listed as
\begin{align}\label{eq:VnNum}
{\cal V}^T \approx 
\left(
\begin{array}{ccccccccc}
0.835 & 0.526 & -0.145 & 0.050 i & 0 & 0 & -0.050 & 0 & 0 \\
-0.247 & 0.600 & 0.761 & 0 & 0.013 i & 0 & 0 & 0.013 & 0 \\
0.488 & -0.601 & 0.633 & 0 & 0 & 0.012 i & 0 & 0 & -0.012 \\
0 & 0 & 0 & 0.707 i & 0 & 0 & -0.707 & 0 & 0 \\
0 & 0 & 0 & 0 & 0.707 i & 0 & 0 & 0.707 & 0 \\
0 & 0 & 0 & 0 & 0 & 0.707 i & 0 & 0 & -0.707 \\
-0.059 & -0.037 & 0.010 & 0.705 i & 0 & 0 & -0.705 & 0 & 0 \\
0.005 & -0.011 & -0.015 & 0 & 0.707 i & 0 & 0 & 0.707 & 0 \\
-0.008 & 0.010 & -0.011 & 0 & 0 & 0.707 i & 0 & 0 & -0.707 \\
\end{array}
\right)
\end{align}
related to the neutrino mass spectrum around $\{0,8\times 10^{-15},5\times 10^{-14},1,1,1,1,1,1 \}$~TeV. And the sneutrino mixing matrix is given numerically by
\begin{align}\label{eq:VsnNum}
{\cal \tilde{V}}^{\cal I} \approx
\left(
\begin{array}{ccccccccc}
0.991 & 0 & 0 & 0.067 & 0 & 0 & -0.118 & 0 & 0 \\
0 & 0.999 & 0 & 0 & 0.018 & 0 & 0 & -0.032 & 0 \\
0 & 0 & -0.999 & 0 & 0 & -0.017 & 0 & 0 & 0.029 \\
-0.131 & 0 & 0 & 0.704 & 0 & 0 & -0.698 & 0 & 0 \\
0 & 0.036 & 0 & 0 & -0.707 & 0 & 0 & 0.706 & 0 \\
0 & 0 & 0.032 & 0 & 0 & -0.707 & 0 & 0 & 0.707 \\
0.037 & 0 & 0 & 0.707 & 0 & 0 & 0.706 & 0 & 0 \\
0 & 0.010 & 0 & 0 & 0.707 & 0 & 0 & 0.707 & 0 \\
0 & 0 & -0.009 & 0 & 0 & -0.707 & 0 & 0 & -0.707 \\
\end{array}
\right)
\end{align}
related to the sneutrino mass spectrum $\{348,349,349,714,708,707,1227,1225,1225\}$~GeV.

Then one can find that all the chargino-sneutrino and the neutralino-slepton diagrams among the non-$\lambda'$
diagrams in the cLFV decays of leptons make negligible contributions due to the vanishing of flavor mixing in sneutrino sector, as shown in Eq.~\eqref{eq:VsnNum}, as well as the diagonal mass matrix of charged slepton for simplicity. As regards $W/H^\pm$-neutrino diagrams, they are always connected to terms ${\cal V}^{T\ast}_{(\alpha+3)v}{\cal V}^{T}_{(\beta+3)v}$, ${\cal V}^{T\ast}_{(\alpha+3)v}{\cal V}^{T}_{\beta v}$, ${\cal V}^{T\ast}_{\alpha v}{\cal V}^{T}_{\beta v}$, and their conjugate terms ($\alpha,\beta=e,\mu,\tau$ and $\alpha\neq\beta$). Readers can see the  calculations of these diagrams in Ref.~\cite{Abada:2014kba}. 
With the numerical form of Eq.~\eqref{eq:VnNum}, the terms ${\cal V}^{T\ast}_{(\alpha+3)v}{\cal V}^{T}_{(\beta+3)v}$ and ${\cal V}^{T\ast}_{(\alpha+3)v}{\cal V}^{T}_{\beta v}$ vanish. The term ${\cal V}^{T\ast}_{\alpha v}{\cal V}^{T}_{\beta v}$ can be decomposed into two parts, $\sum_{N=4}^{9} {\cal V}^{T\ast}_{\alpha N}{\cal V}^{T}_{\beta N}$ and $\sum_{i=1}^{3} {\cal V}^{T\ast}_{\alpha i}{\cal V}^{T}_{\beta i}=-\sum_{N=4}^{9} {\cal V}^{T\ast}_{\alpha N}{\cal V}^{T}_{\beta N}$, related to the nearly degenerate heavy neutrinos and light neutrinos, respectively~\cite{Chang:2017qgi}. Then one can also find that ${\cal V}^{T\ast}_{\alpha v}{\cal V}^{T}_{\beta v}$ makes no effective contribution to the cLFV decays. Thus, we conclude that the non-$\lambda'$ diagrams provide negligible effects on 
the cLFV decays mentioned in section~\ref{sec:loopconstraint} in our input sets.

\bibliographystyle{JHEP}
%\bibliography{ref,ref_prd,references}
\bibliography{ref}

\end{document}